\documentclass[onecolumn]{aa}  
\usepackage{graphicx,color}
\usepackage{txfonts}
\usepackage[colorlinks=true, citecolor=blue]{hyperref}

\begin{document}

   \title{CAPOS: The bulge Cluster APOgee Survey \\   I.  Overview and initial ASPCAP results}

   \titlerunning{CAPOS I}
   \author{Doug Geisler \inst{1,2,3}, Sandro Villanova \inst{1}, Julia E. O\rq{}Connell \inst{1}, Roger E. Cohen \inst{4},
    Christian Moni Bidin \inst{5},
   Jos\'e G. Fern\'andez-Trincado \inst{6}, Cesar Mu\~noz \inst{1,2,3},
   Dante Minniti \inst{7,8,9},
   Manuela Zoccali \inst{7,10}, Alvaro Rojas-Arriagada \inst{7,10}, Rodrigo Contreras Ramos\inst{7,10},  
   M\'arcio Catelan \inst{7,10,11}, Francesco Mauro \inst{5},
   Crist\'ian Cort\'es \inst{7,12}, C. E. Ferreira Lopes \inst{13},
   Anke Arentsen \inst{14}, Else Starkenburg \inst{15,16}, Nicolas F. Martin \inst{14,17}, Baitian Tang \inst{18},
   Celeste Parisi \inst{19,20}, Javier Alonso-Garc\'{i}a \inst{21,7}, 
   Felipe Gran \inst{10,7,22}, Katia Cunha \inst{23,24}, Verne Smith \inst{25}, 
   Steven R. Majewski \inst{26}, 
   Henrik J\"onsson \inst{27},
   D. A. Garc\'ia-Hern\'andez \inst{28,29},
   Danny Horta \inst{30},
   Szabolcs~M{\'e}sz{\'a}ros \inst{31,32},
   Lorenzo Monaco \inst{8}, Antonela Monachesi \inst{2,3},
   Ricardo R. Mu\~noz \inst{33},
  Joel Brownstein \inst{34}, 
  Timothy C. Beers \inst{35},
  Richard R. Lane \inst{6},
  Beatriz Barbuy \inst{36},
  Jennifer Sobeck \inst{37},
  Lady Henao \inst{1},
  Danilo Gonz\'alez-D\'iaz \inst{5,38},
  Ra\'ul E.~Miranda \inst{5},
  Yared Reinarz \inst{5},
  Tatiana A.~Santander \inst{5}
}
   
   \authorrunning{Geisler et al.}

 \institute{Departamento de Astronom\'{i}a, Casilla 160-C, Universidad de
  Concepci\'{o}n, Concepci\'{o}n, Chile
 \and Instituto de Investigación Multidisciplinario en Ciencia y Tecnología, 
Universidad de La Serena. Avenida Raúl Bitrán S/N, La Serena, Chile 
\and  Departamento de Astronomía, Facultad de Ciencias, Universidad de La Serena. Av. Juan Cisternas 1200, La Serena, Chile 
\and Space Telescope Science Institute, 3700 San Martin Drive, Baltimore, MD 21218, USA
\and Instituto de Astronom\'ia, Universidad Cat\'olica del Norte, Av. Angamos 0610, Antofagasta, Chile
\and Instituto de Astronom\'ia y Ciencias Planetarias, Universidad de Atacama, Copayapu 485, Copiap\'o, Chile
\and Millennium Institute of Astrophysics, Santiago, Chile
\and Departamento de Ciencias Fisicas, Facultad de Ciencias Exactas, Universidad Andres Bello, Fernandez Concha 700, Las Condes, Santiago, Chile
\and Vatican Observatory, V00120 Vatican City State, Italy
\and Instituto de Astrof\'isica, Pontificia Universidad Cat\'olica de Chile, Av. Vicu\~{n}a Mackenna 4860, 7820436 Macul, Santiago, Chile
\and Centro de Astro-Ingenier\'{i}a, Pontificia Universidad Cat\'olica de Chile, Av. Vicu\~{n}a Mackenna 4860, 7820436 Macul, Santiago, Chile
\and Departamento de F\'{i}sica, Facultad de Ciencias B\'asicas, Universidad Metropolitana de la Educaci\'on, Av. Jos\'e Pedro Alessandri 774,7760197, Nu\~noa, Santiago, Chile
\and National Institute For Space Research (INPE/MCTI), Av. dos Astronautas, 1758 – São José dos Campos – SP, 12227-010, Brazil
\and Universit\'e de Strasbourg, CNRS, Observatoire astronomique de Strasbourg, UMR 7550, F-67000 Strasbourg, France
\and Leibniz-Institut f\"ur Astrophysik Potsdam (AIP), An der Sternwarte 16, D-14482 Potsdam, Germany
\and Kapteyn Astronomical Institute, University of Groningen, Landleven 12, 9747 AD Groningen, The Netherlands
\and Max-Planck-Institut f\"ur Astronomie, K\"onigstuhl 17, D-69117 Heidelberg, Germany
\and School of Physics and Astronomy, Sun Yat-sen University, Zhuhai 519082, People's Republic of China
\and Observatorio Astron\'omico, Universidad Nacional de C\'ordoba, Laprida 854, X5000BGR, C\'ordoba, Argentina
\and Instituto de Astronom{\'\i}a Te\'orica y Experimental (CONICET-UNC), Laprida 854, X5000BGR, C\'ordoba, Argentina
\and Centro de Astronom\'{i}a (CITEVA), Universidad de Antofagasta, Av. Angamos 601, Antofagasta, Chile
\and ESO Vitacura, Alonso de C\'ordova 3107, Santiago, Chile
\and Steward Observatory, The University of Arizona, 933 North Cherry Avenue, Tucson, AZ 85721-0065, USA
\and Observat\'orio Nacional, Rua General Jos\'e Cristino, 77, 20921-400 S\~ao Crist\'ov\~ao, Rio de Janeiro, RJ, Brazil
\and National Optical Astronomy Observatory, 950 North Cherry Avenue, Tucson, AZ 85719, USA
\and Department of Astronomy, University of Virginia, Charlottesville, VA, 22904, USA
\and Materials Science and Applied Mathematics,
Malm\"o University,
SE-205 06
Malm\"o, Sweden
\and Instituto de Astrof\'isica de Canarias (IAC), E-38205 La Laguna, Tenerife, Spain
\and Universidad de La Laguna (ULL), Departamento de Astrof\'isica, 38206 La Laguna, Tenerife, Spain
\and Astrophysics Research Institute, Liverpool John Moores University, 146 Brownlow Hill, Liverpool L3 5RF, UK
\and ELTE E\"otv\"os Lor\'and University, Gothard Astrophysical Observatory, 9700 Szombathely, Szent Imre H. st. 112, Hungary
\and MTA-ELTE Exoplanet Research Group
\and Departamento de Astronom{\'\i}a, Universidad de Chile, Camino del Observatorio 1515, Las Condes, Santiago, Chile
\and Department of Physics and Astronomy, University of Utah, 115 S. 1400 E., Salt Lake City, UT 84112, USA
\and Department of Physics and JINA Center for the Evolution of the Elements, University of Notre Dame, Notre Dame, IN 46556, USA 
\and Universidade de S\~ao Paulo, IAG, Rua do Mat\~ao 1226, Cidade Universit\'aria, S\~ao Paulo 05508-900, Brazil
\and Department of Astronomy, University of Washington, Seattle, WA, 98195, USA
\and Instituto de F\'isica, Universidad de Antioquia, Calle 70 52-21, Medell\'in, Colombia
}


 
  \abstract
   {Bulge globular clusters (BGCs)
are exceptional tracers of the formation and chemodynamical evolution of this oldest Galactic component.
However, until now, observational difficulties have prevented us from taking full advantage of these powerful Galactic archeological tools.
}
   {CAPOS, the bulge Cluster APOgee Survey, addresses this key topic by observing a large number of BGCs, most of which have only been poorly studied previously. Even their most basic parameters, such as metallicity, [$\alpha$/Fe], and radial velocity, are generally very uncertain. We aim to obtain accurate mean values for these parameters, as well as abundances for a number of other elements, 
and explore multiple populations. In this first paper, we describe the CAPOS project and present initial results for seven BGCs.}
   {CAPOS uses  the  APOGEE-2S   spectrograph observing in the H band to penetrate obscuring dust toward the bulge. For this initial paper, we use abundances derived from ASPCAP,
   the APOGEE pipeline.}
   {We derive mean [Fe/H] values of 
-0.85$\pm${ 0.04} (Terzan 2), 
-1.40$\pm${ 0.05} (Terzan 4), 
-1.20$\pm${ 0.10} (HP 1), 
-1.40$\pm${ 0.07} (Terzan 9), 
-1.07$\pm${ 0.09} (Djorg 2), 
-1.06$\pm${ 0.06} (NGC 6540), and 
-1.11$\pm${ 0.04} (NGC 6642) from three to ten stars per cluster. 
We determine mean abundances for eleven other elements plus the mean [$\alpha$/Fe]
and radial velocity.  CAPOS clusters significantly increase the sample of well-studied Main Bulge globular clusters (GCs) and also extend them to lower metallicity.
We reinforce the finding 
that Main Bulge and Main Disk GCs, formed in situ, have [Si/Fe] abundances slightly higher than their accreted counterparts at the same metallicity.  We investigate multiple populations and find our clusters
generally follow the light-element (anti)correlation trends of previous studies of GCs of similar metallicity. 
We finally explore the abundances of the 
iron-peak elements Mn and Ni and compare their
trends with 
field  populations. 
}
   {CAPOS is proving to be 
   an unprecedented resource for greatly improving our knowledge of the formation and evolution of BGCs and the bulge itself.}


   \maketitle
   
   \titlerunning{test}
   \authorrunning{test}
%

\section{Introduction}

A major goal of modern astronomy is to obtain an understanding of galaxy formation. An ideal tool for this would be a witness  that was both present at the long-since-vanished epoch when most galaxies formed and yet still survives today to tell us its story. We would also like many such witnesses, to corroborate their stories, that are readily observable and yield such critical information as composition, kinematics, and age in a  well-understood, easily measured, accurate and precise way. Enter the globular clusters (GCs). They fulfill all of these attributes admirably and are among our most powerful cosmological archeology probes. 

Globular clusters have proven to be especially vital in piecing together the mass assembly history of our own Galaxy, in particular the halo. The combination of {\em Gaia} astrometry with the above astrophysical data for GCs has recently added a new dimension to our ability to trace back the story of how the Galaxy formed. Armed  with the information provided by their integrals of motion, together with age and metallicity, we can now classify GCs as in situ or accreted, and even distinguish several major accretion events \citep[e.g.,][]{Massari2019, Bajkova2020, Kruijssen2020}. The age-metallicity relationships of the GCs identified with the different proposed progenitors are tight and distinct, indicating that they were indeed likely independent entities that experienced  unique chemical evolution histories \citep{Kruijssen2019, Massari2019, Myeong2019}. Additional clues from detailed abundances of key elements can further constrain our understanding of their particular chemical evolution histories \citep{Horta2020}.

However, such powerful studies require as complete a sample with as detailed knowledge as possible; without this, our ability to distinguish crucial complexities is compromised. This is graphically illustrated by comparing our vast knowledge of halo GCs (and thus the halo and its formation) with our very limited knowledge of their bulge and disk counterparts (and their respective Galactic components). 
We now believe most halo GCs were accreted, while the bulge and disk GCs were formed in situ \citep{Barbuy2018, Massari2019}. The relative ease with which halo GCs can be, and have long been, observed compared to their bulge and disk counterparts means we have learned a tremendous amount about halo formation from them; unfortunately, this is not the case for the disk, or, especially, bulge GCs because of the observational difficulties detailed below. Thus, we face the embarrassing situation that we now know more about the accreted structures than our own homegrown GCs and the story of the main proto-Galaxy's growth they trace. 

The Galactic bulge (GB)  is a fundamental part of our Galaxy. It is one of the most massive Milky Way components and is directly linked to the formation of various other Galactic structures, such as the bar and inner disk. Because galaxies formed from the inside-out, the metal-poor population in the inner few kiloparsecs (kpc) of the GB is the best place to search for the oldest Galactic stars  \citep{Tumlinson2010, Carollo2016, Fragkoudi2020, Horta2021}. Indeed, the rapid chemical evolution in the deep potential well of the GB means that even relatively metal-rich stars could be very old, potentially older than their even more metal-poor halo cousins, which were born and raised in much less massive satellites (Cescutti et al. 2008, Savino et al. 2020).

\begin{figure*}
\centering
   \includegraphics[width=17cm]{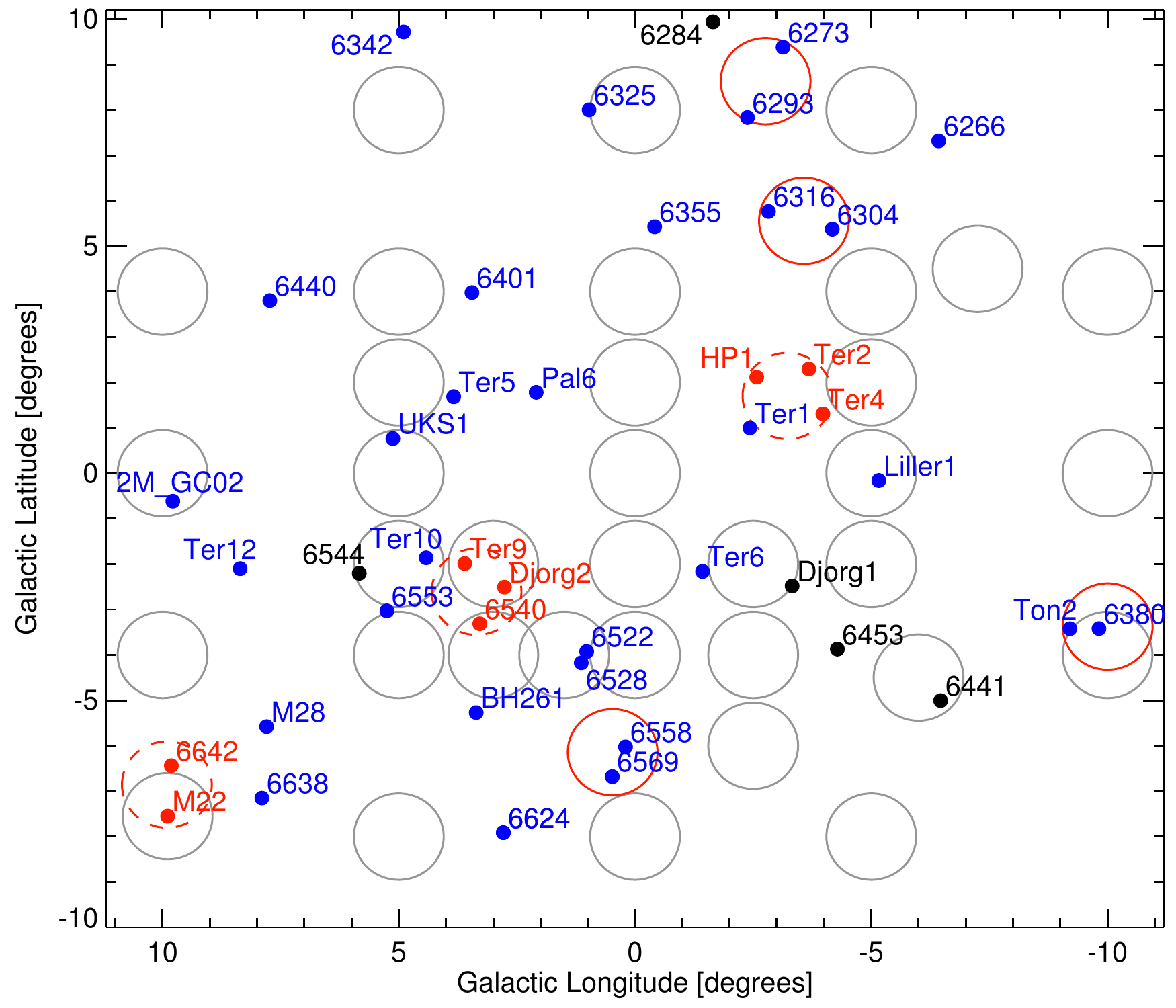}
      \caption{Central $\pm 10^\circ \times \pm 10^\circ $around the Galactic center. BGCs (i.e.,
with $R_{GC}\leq 3.5$kpc) are identified (generally with their NGC number) in blue and red, while other GCs {with $R_{GC}> 3.5$ kpc} are labeled in black.  Gray circles show the locations of the SDSS-IV APOGEE-2S survey fields, while
CAPOS fields are shown in red, with the dashed circles illustrating the CAPOS fields presented here. All CAPOS clusters are BGCs except for M22, which was observed simultaneously with the BGC NGC 6642. A final CAPOS BGC, NGC 6717, is off the plot at ($12.9^\circ, -10.9^\circ$).}
    \label{fields}
\end{figure*}

Bulge GCs (BGCs) are particularly important -- as Shapley famously realized, the GB contains a disproportionately large number of them (43 in the central $\pm 10^\circ \times \pm 10^\circ $ area - Figure\ref{fields}) \citep[][2010 edition; hereafter H10]{Harris1996}. We now know that the GB possesses a GC system, independent from that of the halo \citep{Minniti1995a}, that was formed generally in situ \citep{Tissera2017, Massari2019}. Given our ability to derive absolute GC ages  accurate to the $\sim 0.5$~Gyr level \citep[e.g.,][]{Saracino2016}, depending on the method employed \citep[see][for a recent review]{Catelan2018}, the BGCs very likely include the oldest object in the Galaxy for which we can obtain an accurate age \citep{Dias2016, Barbuy2018, Kerber2019}. This will allow us (once we identify and measure it) to strongly constrain such profound questions as how long after the Big Bang the Galaxy began to form and how.

Unfortunately, until recently we had not been able to unleash the full power of the BGCs to help unravel the history of the GB. Despite its proximity and central role as a primary primordial portion of the Galaxy, the GB has resisted detailed investigation due to the high foreground extinction that strongly limits optical observations. An additional complicating factor is  crowding, which prohibits accurate measurements of individual stars. Field contamination is also highly problematic as what we see as the GB in projection is really the superposition of not only the bulge, but also the bar, the foreground (and background) thin (and thick) disk(s) and the halo. Furthermore, BGCs are generally buried in densely populated fields whose stars are difficult to distinguish photometrically from those of the BGC we want to study. Consequently, the two most comprehensive {\em Hubble Space Telescope (HST)} surveys of GCs - the {\em Advanced Camera for Surveys (ACS)} survey \citep{Sarajedini2007} and the HST UV Legacy Survey (Piotto et al. 2015) -  were focused on halo and disk objects, almost totally avoiding those in the bulge. 

However, the first two problematic effects are minimized by observing in the near-IR with high spatial resolution detectors, which have, fortunately, recently come on line. Moreover, contamination is now greatly alleviated via Gaia astrometry \citep{Gaia-Collaboration2016, Gaia-Collaboration2018}. Also, high-precision radial velocities such as those provided by the {\em Apache Point Observatory Galactic Evolution Experiment (APOGEE)} \citep{Majewski2017} bring an additional powerful membership discriminant. The combination of all these recent 
advances finally enables us to effectively exploit the extraordinary Galactic archeology attributes of BGCs. 

Recognizing the importance of near-IR observations of the GB, the Vista Variables in the Via Lactea (VVV) survey \citep{Minniti2010} was granted
a 5 year ESO Public Survey on the ESO 4m VISTA telescope to map the GB (and adjacent disk) in $YZJHK_s$ with the aim of
studying its structure and a vast variety of fascinating objects hidden there, with BGCs as one of the top priorities. The VVV has
proven to be extremely successful, so much so that a 3 year extension, dubbed the VVV eXtended survey, or VVVX for short, is now underway
\citep{Minniti2018}. The VVVX extends the spatial coverage of near-IR imaging of
the bulge out to $10^\circ$ in both Galactic latitude and longitude - the entire area of Fig.\ref{fields}.

The VVV team has taken full advantage of this unique database  to study BGCs. They have discovered many new GC candidates \citep[e.g.,][]{MoniBidin2011, Minniti2011, Minniti2017a, Minniti2017b, Palma2019}, found dual horizontal branches in two metal-rich GCs \citep{Mauro2012}, investigated structural parameters  \citep{Cohen2014}, studied variable stars \citep[e.g.,][]{Alonso-Garcia2015}, and provided the deepest VVV-based color-magnitude diagrams (CMDs) to date \citep{Cohen2017}. In addition, and of
particular importance, follow-up very deep near-IR HST \citep{Cohen2018} and Gemini-S GeMS \citep[e.g.,][]{Saracino2016} images have been obtained for many BGCs.  GeMS is a multiconjugate adaptive optics instrument that produces very high spatial resolution images in the near-IR, with image quality competitive with HST but on an 8m telescope. The superb image quality of both HST and GeMS in the near-IR makes BGCs accessible, for the first time, to the deep photometry required to derive accurate ages. The VVV images, as pioneering as they are, are simply too shallow and the spatial resolution is woefully inadequate for this purpose. For example, the depth of the exquisite GeMS NGC 6624 data \citep{Saracino2016}, reaching >4 mag below the main sequence turn-off (MSTO) in $K_s$ and >3 mag fainter than VVV, reveals faint features previously only theoretically predicted, such as the main sequence (MS) knee - the blueward inflection in the curvature of the lower MS - allowing age measurements with 0.5 Gyr precision from purely near-IR ground-based imaging, even in the cluster center.  Similar deep GeMS and HST optical-near-IR images for a number of BGCs are now in hand, providing an unprecedented photometric database for a large number of BGCs. 

However, to measure the best age, and hence pin down the earliest formation epoch of the Galaxy with the smallest error, requires both deep high spatial resolution photometry and high spectral resolution spectroscopy to derive the detailed abundances required for isochrone fitting. The main factors that are still sorely lacking are good [Fe/H], $[\alpha$/Fe], and [CNO/Fe] values for BGCs, all of which are very scarce. All CMD-based age diagnostics are very sensitive to these key elements. Given our exquisite near-IR photometry exemplified by NGC 6624 (also see, e.g., HP1 - \citet{Kerber2019}, NGC 6256 - \citet{Cadelano2020}), the single remaining dominant uncertainty on BGC ages is their abundances. Incredibly, most of these invaluable objects have only poor estimates of the overall metallicity - based on a variety of techniques, making a very heterogeneous sample - and little to no abundance information beyond this.

Fortunately, with the arrival of the APOGEE southern instrument APOGEE-2S, filling in the missing link of abundances in BGCs is now feasible. APOGEE is a high-resolution, near-IR spectroscopic survey, which is part of the fourth iteration of the Sloan Digital Sky Survey \citep[SDSS-IV;][]{Blanton2017}. APOGEE-2S, a copy of the original APOGEE instrument built for the Sloan telescope in the north for SDSS-III, is attached to the du Pont 2.5m telescope at Las Campanas Observatory in Chile, opening access to the Southern Hemisphere and the GB as part of SDSS-IV. 

Given their importance, the SDSS-IV survey has targeted some BGCs with APOGEE-2S. Nevertheless, this is an enormous global Galactic and even extragalactic survey, covering hundreds of thousands of stars, and simply cannot afford to study a single part of the Galaxy in complete detail.  SDSS-IV is surveying 35 fields within the  central $10^\circ \times 10^\circ$ GB area on the sky. These fields include a number of GCs. 
However, the GB is more than just an area on the sky:  It is a distinct Galactic component with a well-known spatial distribution, and not all of the objects found in this area  are actually inside the GB - some are in the foreground or background. Taking 3.5 kpc as a limiting Galactocentric radius for the GB \citep{Schultheis2017, Massari2019}, the H10 catalog of GC properties includes 55 BGCs. We note that \citet{Bica2016} use both the spatial distribution, with a 3 kpc cutoff, as well as metallicity to arrive at a final list of 43 BGCs. For the present, we only concern ourselves with the three-dimensional spatial location, using a 3.5 kpc limit,  and ignore metallicity, orbit, and origin when refering to BGCs. Of the 43 GCs lying in the central $10^\circ \times 10^\circ$ (which includes most but not all BGCs), 6 lie well beyond
this Galactocentric radius, including 4 of the SDSS-IV GCs, leaving a total of 37 bona fide BGCs.
SDSS will only observe a handful of these. Initial small-scale studies of APOGEE results based on SDSS-III and -IV observations are given in \citet{Schiavon2017}, \citet{Tang2017}, and \citet{Fernandez-Trincad2019}. Indeed, \citet{Meszaros2020} present the APOGEE sample of 44 clusters, of which only 8 are bona fide BGCs according to \cite{Massari2019}. 
Of these, they dismiss all but 2 as either not having a large enough sample of well-observed members or having too high reddening. This represents less than 4\% of the total number of BGCs known, a disconcertingly low value. If one is searching for the oldest GC in the Galaxy, and aiming to take full advantage of the wealth of astrophysical detail these key objects can provide by performing a definitive study of the  BGC system, as complete a sample as possible is essential.

Hence the CAPOS project, the bulge Cluster APOgee Survey (the b is silent). 
The primary goal of CAPOS is to obtain detailed abundances and kinematics for as complete a sample as possible of bona fide members in true BGCs,  using the unique advantages of APOGEE to complement the much smaller sample observed by SDSS-IV, via Chilean access to APOGEE-2S through the CNTAC (the Chilean National Telescope Allocation Committee).
We aim to help gather the first definitive database on the BGC system, search for
the oldest object in the Galaxy with an accurate age, study the very complex nature of the GB and uncover any underlying correlations, determine BGC velocities and orbits and investigate multiple populations in the BGC system, and compare them to their halo (and disk) counterparts. As noted above, this demands as complete a  sample as possible. The power of this approach is to establish the population of BGCs as an ensemble, finally placing them on a level with and in the context of their hither-to much better-studied halo counterparts, and thus fill in our picture of all Milky Way GC systems, especially the in situ BGCs.

CAPOS originally targetted 20 cataloged BGCs, with the goal of bringing the number of bona fide BGCs observed by APOGEE to 28, over half of the total, with CAPOS supplying almost three-fourths of the observed sample. CAPOS will, together with the SDSS-IV data, provide a legacy database of the  BGC system. This will provide much better and completely self-consistent spectroscopic metallicities than currently available for all the observed BGCs, as well as derive detailed abundances for some 20 elements with a wide variety of nucleosynthetic origins, precise to $\sim 0.05$ dex. Current knowledge of their abundances is limited to only a few, if any, spectra of individual stars, and/or crude photometric metallicity estimates, or even integrated light measurements in many cases. We will
also obtain excellent radial velocities. Many of the BGCs have only limited or even nonexistent velocity information.

CAPOS also targets any recently discovered BGC candidates that can be observed serendipitously  with cataloged BGCs, given the large field of view of APOGEE. These include such intriguing objects as FSR1758, a newly discovered massive BGC that is the eponymous member of the Sequoia dwarf galaxy  \citep{Barba2019, Massari2019, Myeong2019, Villanova2019, Romero-Colmenares2021}, as well as any other candidates from the VVV survey \citep[e.g.,][]{Palma2019}.

We will also search for multiple populations (MP) in BGCs. The study of MP in GCs has revolutionized our understanding of their formation and evolution but so far has been limited almost exclusively to non-GB, non-metal-rich GCs. CAPOS will open up the regime of high metallicity to detailed MP studies, since BGCs include the highest metallicity of all Milky Way GCs. APOGEE includes lines of the light elements C, N, O, Na, Mg, Al, and Si, which are essential to tracing MP. 

We aim to observe  5-10 members per cluster to derive accurate mean abundances, search for and characterize MP, and constrain scenarios for the formation of MP.

Given the compact size of BGCs and fiber collision limitations, most fibers will be available for objects outside of the GCs. To optimize the science return, CAPOS targets
hundreds of GB field giants per plate. Abundances for Fe and $\alpha$-elements and velocities will delineate the kinematics and chemistry of distinct GB components. Our fields also overlap K2 mission areas \citep{Howell2014}, and we will exploit K2 data to explore Galactic archaeology using asteroseismology \citep{Johnson2016}. Gaia astrometry, K2 asteroseismology, CAPOS stellar parameters, and VVV photometry allow us to trace the GB chemical evolution, resolved into its different components.

Finally, we also target metal-poor candidates from the {\em Extremely Metal-poor BuLge stars with AAOmega survey (EMBLA)} \citep{Howes2016} and {\em Pristine Inner Galaxy Survey (PIGS)} surveys \citep{Starkenburg2017a, Arentsen2020}.
The metal-poor tail of the bulge/inner Galaxy has not yet been studied in detail since the number of metal-poor stars is extremely small compared to the more metal-rich stars that dominate in the inner Galaxy. In APOGEE DR16, there are only $\sim 50$ stars with [M/H] $< -2.0$ located within $(|l, b|) < 10^\circ$ \citep{DR16}. Past and current high-resolution surveys targeting metal-poor stars  only contain a handful  with $-2.5 <$ [Fe/H] $< -1.5$ \citep{Howes2016, Duong2019, Lucey2019}. Larger samples of metal-poor inner Galaxy stars are needed to disentangle this complicated area of the Galaxy, where multiple Galactic components overlap. Additionally, the contribution of disrupted GCs to the metal-poor inner Galaxy is currently poorly constrained. It is crucial to obtain high-resolution follow-up spectra for metal-poor inner Galaxy stars, providing detailed chemical abundances combined with kinematics, which can help to disentangle different stellar populations. Finally, it is also of great interest to search for the most metal-poor stars in the inner Galaxy, which are likely to be among the oldest in the Milky Way \citep[e.g.,][]{Tumlinson2010, Starkenburg2017b, Horta2021}.

The CAPOS/PIGS  collaboration aims to greatly increase the number of (very) metal-poor inner Galaxy stars with high-resolution spectroscopy available. The Pristine survey \citep{Starkenburg2017b} is a photometric survey that employs metallicity-sensitive CaHK photometry from MegaCam on the CFHT to efficiently search for the most metal-poor stars in the Galaxy. Its main focus is on the Galactic halo and dwarf galaxies; additionally, there is a sub-survey toward the GB (PIGS). 
There are two CAPOS fields with PIGS targets.
Based on the results from the first field, we expect to increase the number of [M/H] $< -2.0$ stars within $(|l, b|) < 10^\circ$ by 60\% compared to APOGEE DR16, and we expect to increase the number of stars with [M/H] $< -1.5$ by 30\%. 
Most of these stars will also have low/intermediate resolution optical/calcium triplet spectra available from the PIGS follow-up efforts \citep{Arentsen2020}. 

In this first CAPOS paper, we include an overview of the project and initial results based on the {\em APOGEE Stellar Parameters and Chemical Abundance Pipeline  - ASPCAP} \citep{GarciaPerez2016} analysis of all BGC CAPOS data released in DR16 \citep{DR16}. As can be seen in Fig.\ref{fields}, the other H10 BGCs CAPOS observed include:
NGC 6273, NGC 6293, NGC 6304, NGC 6316, 
NGC 6380, Ton 2, 
NGC 6558, NGC 6569, 
and NGC 6717 (this final BGC is outside Fig.\ref{fields}).
The paper is organized as follows: We first present details of the selection of clusters as well as targets within clusters. We also discuss how field stars were selected, including bulge field stars, K2 stars, and stars from the EMBLA and PIGS surveys. Next, the observations and reductions are discussed. We then present a number of key results from the ASPCAP analysis, starting with the final determination of cluster members based on a variety of criteria. We discuss the ASPCAP atmospheric parameters and their errors. Next we derive the mean metallicity, $[\alpha/{\rm Fe}]$, and heliocentric radial velocity of each cluster, and compare these values to the literature. We then investigate mean abundances for a number of other well-determined elements. Multiple populations in the clusters are then investigated, followed by the results for Fe-peak elements. Finally, we summarize our main conclusions.

\section{Sample selection}

\subsection{Cluster selection}

The initial CAPOS goal was to observe {\it all} BGCs that appear in the H10 catalog and lie within the central $10^\circ \times 10^\circ$  area of the bulge that were not planned to be observed with APOGEE-2S as part of the SDSS-IV survey. This would have provided a complete sample of H10 BGCs within this central bulge area, and would have included, together with the SDSS clusters, some 70\% of the cataloged BGCs. However, this proved to be too ambitious, given the time constraints imposed by the limitations on the instrument,  Chilean access, and the  extensive Covid-19 LCO shutdown coming in the last year of APOGEE-2. In the end, we were able to observe a total of 16 CAPOS BGCs. This, combined with the SDSS clusters, will bring the total to 24 of 55 BGCs,  a nearly majority sample that doubles the total covered by APOGEE over those observed by SDSS alone. In addition, we will observe a few BGCs or candidates that are not in the H10 catalog that lie within the same APOGEE field as cataloged targets.

Specific APOGEE fields to observe were selected via several criteria. First, given the limited time granted, we prioritized fields that included multiple BGCs within the same large APOGEE field of view, which is over $1^\circ$ in radius. We also prioritized the most metal-poor BGCs, given our goal of searching for the oldest of these. Everything else being equal, the least-studied clusters were deemed especially interesting. Fields that included new GC candidates \citep{Minniti2017a, Minniti2017b, Barba2019, Palma2019} were also prioritized. Finally, due to the hour angle restrictions of APOGEE,
especially during the highly competitive bulge season, we emphasized fields at the extreme RA ends of the bulge. The initial three fields observed and reported on here include two with three BGCs each and a third with a BGC and a non-BGC. This last field is at the extreme eastern end of the central bulge field, and includes NGC 6656 (M22). Although M22 is outside of our limiting Galactocentric radius for BGCs, it is an interesting GC in its own right, having been the subject of considerable debate as to whether or not it has an intrinsic metallicity spread \citep[e.g.,][]{Norris1983, Mucciarelli2015}, and is readily observed simultaneously with the BGC NGC 6642. M22 will be the subject of another paper in this series.
Finally, the GC candidate Minni 51 \citep{Minniti2017a} was observed along with 3 BGCs in one of our fields.  We note that our CAPOS observations find no convincing evidence for the reality of Minni 51, i.e. there is no clustering in metallicity:radial velocity space for the 9 targets, supporting the null finding by \citet{Gran2019}, and we will not discuss this object further here.

We note that all of the targets designated BGC here are also classified as bona fide BGCs by \citet{Bica2016}, based on their spatial distribution and  metallicity, and as Main Bulge objects by \citet{Massari2019} as judged by their kinematics. \citet{Perez-Villegas2020} also identify all of our clusters as bulge/bar from their orbits. The recent reassessment by \citet{Bajkova2020} maintains the same association as Massari for our objects except for NGC 6540, which they label a Disk GC. 
We will keep the Massari designations but note that CAPOS velocities will be used to reinvestigate in detail this association in a forthcoming paper.
For completeness sake, we report in Table 1 basic positional data for all CAPOS clusters, including the APOGEE field ID, while Table 2 lists additional basic parameters from the literature for the clusters we report on here. We note that our BGC sample covers a wide metallicity range, but not the very highest BGC metallicities, which reach nearly solar abundance, and that the reddenings are generally quite large.

\begin{table*}
	\begin{center}
		\setlength{\tabcolsep}{2.mm}  
		\caption{Basic positional data for all CAPOS clusters.}
		\begin{tabular}{c c c c c c }
			\hline
			Cluster ID & $\alpha$ (J2000.0) & $\delta$ (J2000) & L($^\circ $) & B($^\circ $) & APOGEE  \\
			&   $hh:mm:ss$ & $^\circ $~:~ $ ' $~: ~ $''$  &  & & Field  \\
			\hline
NGC 6273 & 17 02 37.8 & -26 16 04.7  & 356.87  &  9.38& 357+09-C \\
NGC 6293  &  17 10 10.2 & -26 34 55.5  & 357.62  &  7.83 & 357+09-C \\
NGC 6304 & 17 14 32.3 &  -29 27 43.3  & 355.83  &  5.38 & 356+06-C \\
NGC 6316 & 17 16 37.3 &  -28 08 24.4  & 357.18  &  5.76 & 356+06-C \\
Terzan 2 & 17 27 33.1 & $-$30 48 08.4 & 356.32  &  2.30 & 357+02-C \\
Terzan 4 &  17 30 39.0 & $-$31 35 43.9  & 
356.02  &  1.31 & 357+02-C \\
HP1	&  17 31 05.2 & $-$29 58 54 & 
357.44  &  2.12 & 357+02-C \\
NGC 6380 & 17 34 28.0 &  -39 04 09  &   350.18 &  -3.42 & 350-03-C \\
Ton 2 & 17 36 10.5  & -38 33 12     & 350.80  & -3.42 & 350-03-C \\
Terzan 9 &  18 01 38.8 & $-$26 50 23 & 
3.61  & -1.99 & 003-03-C \\
Djorg 2	& 18 01 49.1 & $-$27 49 33 & 
2.77  & -2.50 & 003-03-C \\
NGC 6540	&  18 06 08.6 & $-$27 45 55 & 3.29  & -3.31 & 003-03-C \\
NGC 6558 & 18 10 17.6 &  -31 45 50.0  &   0.20  & -6.02 & 000-06-C \\
NGC 6569 & 18 13 38.8 &  -31 49 36.8  &   0.48  & -6.68 & 000-06-C \\
NGC 6642	& 18 31 54.1 & $-$23 28 30.7 & 9.81 & -6.44 & 010-07-C \\
NGC 6656	&  18 36 23.9 & $-$23 54 17.1 & 9.89 &  -7.55 & 010-07-C \\
NGC 6717  & 18 55 06.0 &  -22 42 05.3  &  12.88 & -10.90 & 013-11-C \\

			\hline
		\end{tabular}  \label{basicparamsall}\\
	\end{center}
	\raggedright{{\bf Note:} Equatorial coordinates, L and B are from H10. }
\end{table*}


\begin{table*}
	\begin{center}
		\setlength{\tabcolsep}{2.mm}  
		\caption{Basic parameters from the literature for CAPOS I clusters.}
		\begin{tabular}{ c c c c c c }
			\hline
			Cluster ID & $\alpha$ (J2000.0) & $\delta$ (J2000) & L($^\circ $) & B($^\circ $) & APOGEE  \\
			&   $hh:mm:ss$ & $^\circ $~:~ $ ' $~: ~ $''$  &  & & Field  \\
			\hline
\hline
Cluster ID & [Fe/H] & E(B-V) & $V_r$ & $\mu _{\alpha} cos\delta$ & $\mu _{\delta}$  \\
&    & & km $s^{-1}$  & mas $yr^{-1}$ & mas $yr^{-1}$ \\

\hline
Terzan 2 &   -0.69 & 1.87 & 129.0$\pm$1.2 & -2.20$\pm$0.10 &  -6.21$\pm$0.09\\
Terzan 4 &  -1.41 & 2.00 & -39.9$\pm$3.8  & -5.36$\pm$0.07 &  -3.35$\pm$0.06\\
HP1	&  -1.00  & 1.12 & 40.6$\pm$1.3 & 2.54$\pm$0.06 &  -10.15$\pm$0.06\\
Terzan 9 &   -1.05  & 1.76 & 29.3$\pm$3.0 & -2.17$\pm$0.06 &  -7.40$\pm$0.05\\
Djorg 2	&  -0.65 & 0.94 & -148.1$\pm$1.4 & 0.54$\pm$0.04 &  -3.04$\pm$0.03\\
NGC 6540	&   -1.35 & 0.66 & -18.0$\pm$0.8 & -3.80$\pm$0.05 &  -2.73$\pm$0.05\\
NGC 6642	&  -1.26 & 0.40 & -33.2$\pm$1.1 & -0.19$\pm$0.03 &  -3.90$\pm$0.03\\
			\hline
		\end{tabular}  \label{basicparams}\\
	\end{center}
	\raggedright{{\bf Note:} [Fe/H] and E(B-V) are from H10. $V_r$ and astrometry are from \citet{Baumgardt2019}. }
\end{table*}

\subsection{Cluster target star selection}

For each cluster, available spectroscopic, photometric, and astrometric information were all leveraged wherever possible to maximize the number of bona fide cluster members that were observed, and exclude disk and bulge field star contaminants.  Specifically, the following criteria were used to assign priorities, from highest to lowest:
\begin{itemize}
    \item Stars with extant high-resolution spectroscopy, noting that only 5 of our 16 CAPOS clusters have published high-resolution spectroscopy, and only two in the current sample. This is to ensure membership and allow comparison with previous studies.

    \item Stars that are members based on radial velocity and metallicity information from low-resolution spectroscopy, typically using the CaII triplet.  Such information was available for 10 of 16  CAPOS clusters, drawn primarily from \citet{Mauro2014}, \citet{Dias2016}, and \citet{Geisler2021}.
    
    \item Stars that are likely members based on their proper motions (PMs) and uncertainties. For clusters observed in the first (2018) season (the current sample), ground-based PMs were employed where available, especially those derived from multi-epoch VVV photometry \citep{ContrerasRamos2017}, since Gaia data were not yet available.
      For the remainder of the clusters, a similar procedure was followed to select members from all candidates with valid PMs from Gaia DR2, typically applying a 2$\sigma$ clip to the two-dimensional proper motion errors of stars in the magnitude range of APOGEE targets (see below) to select candidate members.    
    
    \item Near-IR PSF photometry from the VVV survey \citep{Cohen2017, Alonso-Garcia2018}, matched to 2MASS, was used to reject foreground disk stars via a cut in $(J-K_{S})$ color blueward of the cluster red giant branch
    
\end{itemize}

The last requirement on cluster targets is set by the available exposure time, which was a total on-source time of 5h or 6h per cluster initially.  Following the strategy adopted for the APOGEE survey of GCs \citep{Zasowski2017}, targets were required to have 7.5$< H <$13 to ensure useful signal-to-noise.  In practice, these cuts imply that all BGC members actually observed should be GK giants.  

With a list of candidate members in hand for each cluster, targets were assigned to individual plates, assigning the brightest stars first within each priority itemized above to maximize the total number of candidates observed.
By far the primary factor limiting the number of candidate members that could be observed in each cluster is the 56$\arcsec$ fiber collision limit for the southern APOGEE spectrograph, which is why careful prioritization of candidate members as described above is crucial to maximize the number of likely cluster stars assigned fibers. We employed multiple visits, changing out bright stars in different visits, to observe as many probable cluster members as possible (see Sect. 3). The number of visits was decreased as the survey progressed, but for the current sample, 6 visits were obtained for field 003-03-C and 5 each for fields 357+02-C and 010-07-C.
Both GC and field targets are initially selected in three magnitude ranges: $H\leq11$,
$11\leq H\leq 12.2$, and $12.2< H\leq 12.8$, for single, 2-3 and $>3$ visits, respectively, with the aim of achieving roughly similar S/N for the final added spectra. 

\subsection{Bulge field star selection}
\begin{figure*}
\centering
   \sidecaption
   \includegraphics[width=14cm]{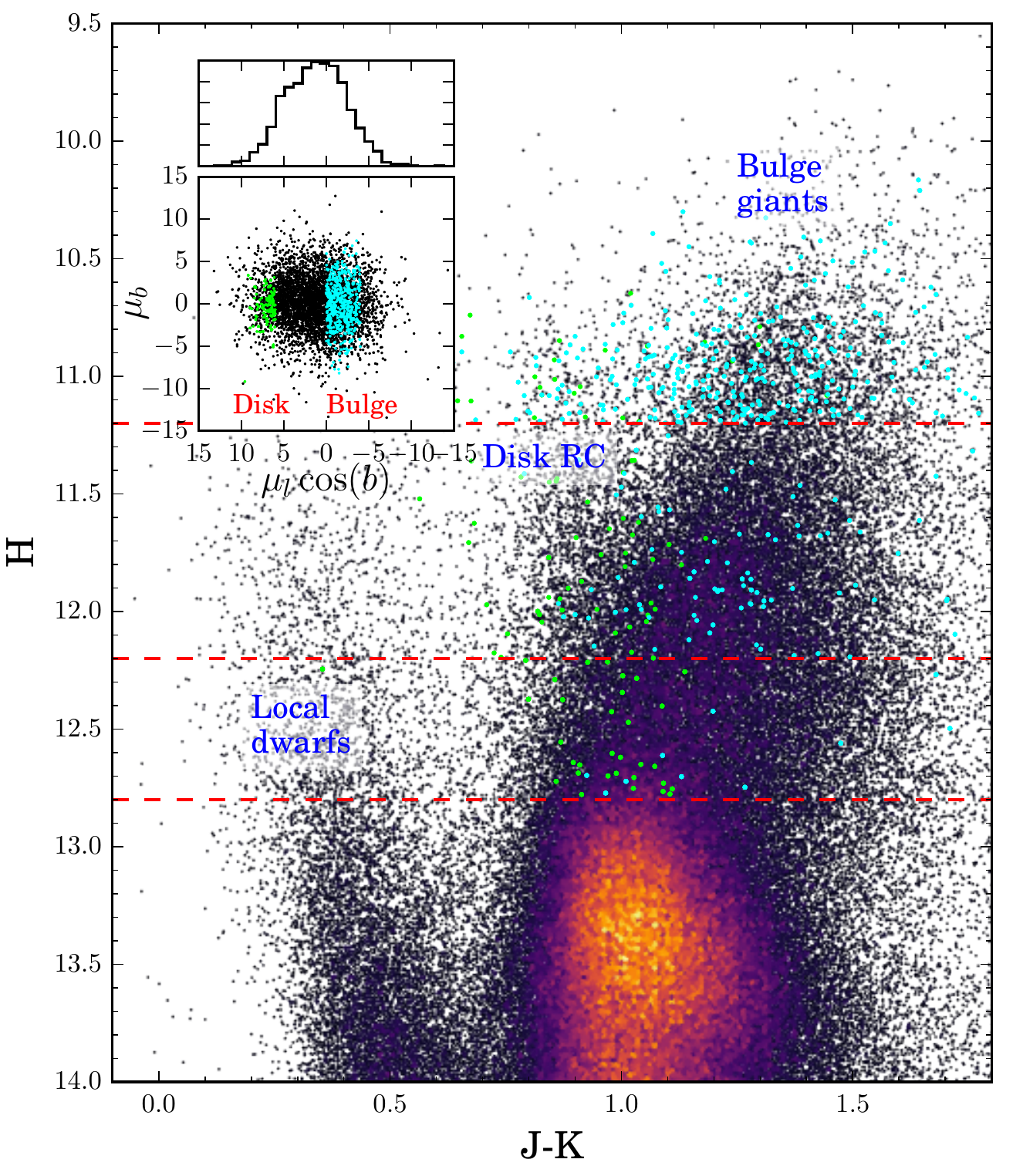}
      \caption{Selection function of bulge and disk field giants. \textit{Main panel:} Annotated CMD of a typical bulge field (003-03-C) is shown as a Hess diagram. The horizontal red dashed lines indicate the magnitude limits adopted to select cohorts, as described in the main text. \textit{Inset panel: }$\mu_l \cos(b)$~vs.~$\mu_b$ VVV PM diagram. The black points correspond to a random 2500 field star subsample with small errors in PMs. The cyan (green) points in both panels display the kinematically selected bulge (disk) giants from the magnitude ranges displayed in the main panel.}
    \label{bulgefield}
\end{figure*}

The APOGEE 2.1$^\circ $ diameter circular FOV covers an area that is not only
large enough to contain more than one GC in some favorable
cases, but also allows us to design
observations that complement GC targets with surrounding field populations. In this regard, 
a selection function was designed to also include bulge and disk field giants, in the fields observed within the initial CAPOS
project, namely fields 010-07-C, 003-03-C, and 357+02-C. 
These fields were observed before the release of Gaia DR2, and thus necessitated a different procedure to include PM criteria than was the case for subsequent observations.
The selection was
designed to observe bulge RGB stars, plus a fraction of disk red clump (RC)
stars. The selection was based on photometry and relative PMs computed
from VVV psf photometry \citep{ContrerasRamos2017}. The selection function is
illustrated in Fig.\ref{bulgefield}. In this figure, the CMD of a typical bulge region
is displayed (in this case, the 003-03-C field). The dominant bulge RC population is prominent around ($J-K=1.1, H=13.5$), from which
the sequence of intrinsically brighter and redder RGB stars emerges. The bright 
vertical plume at J-K=0.4 corresponds to Solar Neighborhood dwarf stars,
while the vertical, less prominent sequence, at J-K=0.8 - 0.9, is due to disk RC stars at
progressively larger distances from the Sun.

Using VVV PMs,
the vector point diagram (VPD) was constructed for the whole set of
available stars in a few magnitude ranges/cohorts (see inset in
Fig.\ref{bulgefield}). From this diagram, two kinematical selections are performed to
obtain clean samples from the bulge RGB and disk RC. This is done by
selecting stars from the bulge and disk overdensities in the VPD, avoiding
the overlap area in between them. Disk stars are selected as those with
$6\leq \mu_l \cos(b)\leq 10$, while bulge stars are those with $-4\leq \mu_l
\cos(b)\leq 0$. The resulting kinematically selected samples are highlighted in the VPD and the CMD displayed in Fig.\ref{bulgefield} as green and cyan dots, respectively. For subsequent observations, we used the same procedure but supplemented with Gaia DR2 PMs.

\subsection{K2 star selection}
Kepler K2 targets were selected from the K2 Galactic Archaeology Program (K2GAP), described by \citet{Stello2015, Stello2017}. Specifically, the K2 sample is composed of  red giants located in the K2 campaigns 7 and 11, which are overlapping with several surveys, including VVV, Gaia, and 2MASS. These stars present $H$ magnitudes ranging from $7.2$ to $12.1$. The asteroseismology observations performed by K2  provide us with accurate measurements of stellar masses, ages,  and radius for red  giants \citep{Kallinger2010}. 
K2 targets appear in the initial CAPOS field 357+02-C, and were given high priority.

\subsection{EMBLA star selection}
The EMBLA survey contains some of the oldest and most metal-poor stars in the bulge and indeed in the Galaxy \citep{Howes2016}. 
Of course, some fraction of the EMBLA stars are metal-rich due to contamination. EMBLA sources overlap spatially with two initial CAPOS fields, 003-03-C
and 010-07-C.
EMBLA targets in these fields were selected and prioritized using a combination of preexisting information and observing constraints imposed by APOGEE-2S magnitude and fiber collision limits \citep{Zasowski2017, Santana2021}. In the selected fields, 
EMBLA sources were given relatively high priority and chosen at random at the fiber configuration stage. 

\subsection{PIGS star selection}

Metal-poor PIGS targets are selected using the metallicity-sensitive Pristine $CaHK$ photometry. The $CaHK$ photometry, which is already cross-matched with \textit{Gaia} DR2 \citep{Gaia-Collaboration2016, Gaia-Collaboration2018}, is also cross-matched with 2MASS \citep{Skrutskie2006}. We select stars with $H < 12.8$ and 2MASS quality $flag ph\_qual = AAA$, to have stars in the CAPOS brightness regime with good quality 2MASS photometry. The Gaia and Pristine $CaHK$ photometry are corrected for extinction using the \citet{Green2018} reddening map. 
We limit the selection to a color range of $0.9 < (BP - RP)_0 < 1.5$, where metal-poor giant stars are expected to lie. A cut is added on the (parallax/parallax\_error) $< 0.3$, to remove contamination from foreground stars with significant parallaxes. We finally sort the stars by the following metallicity-sensitive color: $(CaHK - BP)_0 - 2.5(BP - RP)_0$, and select the 100 stars with the smallest values, which are expected to be the most metal-poor. The next best 200 stars are provided as back-up targets. The efficiency of this type of selection has been presented in  \citet{Arentsen2020}. The CAPOS/PIGS collaboration began with the 2019 observations, so no results are included in this work, which contains only earlier observations.

\section{Observations and reductions}

CAPOS time was granted via the CNTAC as an External Program to APOGEE \citep{Majewski2017}, part of SDSS-IV \citep{Blanton2017}. The APOGEE instruments are high-resolution,
near-infrared H-band spectrographs \citep{Wilson2019} observing from both the Northern Hemisphere at Apache Point Observatory (APO) using the SDSS 2.5m telescope \citep{Gunn2006}, assisted by the NMSU 1m \citep{Holtzman2010}, and the Southern Hemisphere at Las Campanas Observatory using the 2.5m du Pont telescope \citep{Bowen1973}. Stars are targeted using selections described in \citet{Zasowski2013}, \citet{Zasowski2017}, \citet{Santana2021} and \citet{Beaton2021}.
Spectra are reduced as described in \citet{Nidever2015} and analyzed using the APOGEE Stellar Parameters and Chemical Abundance Pipeline \citep[ASPCAP,][]{GarciaPerez2016}, which compares the observed spectra with a  
complete spectral library of synthetic spectra \citep[e.g.,][]{Zamora2015}
A detailed analysis of the
accuracy and precision of the stellar parameters and abundances can be found in \citet{Holtzman2018} and \citet{Henrik2018}. Our analysis uses results from the 16th Data Release (DR16) of the SDSS collaboration \citep{DR16},
which is the first data release containing
data from the southern instrument. Further explanations and assessments of this data release can be found in \citet{Henrik2020}.

CAPOS was awarded time
over a number of different semesters. The assigned CAPOS nights were
June 20-21 (CNTAC program CN2017B~-~37) and July 21, 2018 (CN2018A~-~20), June 14 and 19 (CN2018B~-~46) and July 9-10, 2019 (CN2019A~-~98), and May 30, June 27, and August 10, 2020 (CN2019B~-~31), for a total of 10 nights. Unfortunately, all three 2020 nights were lost to LCO closure due to Covid-19, but an additional BGC was very recently observed after APOGEE-2S operations reinitiated in late October while the bulge was still observable.
No further CAPOS observations will be possible as SDSS-IV has now ceased operations. 
DR16 only includes data taken by June 2019 and thus only includes three CAPOS fields. These three fields include seven BGCs, M22, and a candidate GC from \citet{Minniti2017a}. M22, along with other CAPOS clusters observed subsequently to June 2019,  will be studied in subsequent papers.

CAPOS observations were carried out by the SDSS APOGEE-2S survey team in the same manner as for the survey observations \citep[see][for details]{DR16}, except that CNTAC time was granted as entire nights. Unfortunately, given the rather strict hour angle limits of APOGEE-2S, the bulge is only observable during about half the night, even during bulge season in the long winter nights. This of course severely limits the number of possible CAPOS observations, and caused us to descope the project by observing significantly fewer fields and less time per field than originally planned. Any unusable time during each assigned CAPOS night was returned to SDSS for normal survey operations. 

To obtain as large a sample of cluster members as possible to map potential chemical inhomogeneities within a cluster, in particular multiple populations, our goal is to observe 5-10 members per cluster. However, 
given the small size of
GB clusters  and the fiber collision
limit, we are able to put only a small number of fibers on probable cluster members for each pointing.
To mitigate this problem, CAPOS generally observes each field with multiple
visits (with the standard 1 hour exposure per visit). The number of multiple visits per cluster was originally set to six, but we trimmed this to five after the first run and eventually even lower, given the above time constraints.
The number of targets observed in each cluster is
maximized by replacing bright (H<11) stars with other bright stars after a single visit, (or after three
visits for 11< H <12.2), while fainter (12.2< H <12.8) targets are observed for all visits to attain
the required S/N ($\ge 70$ is the standard minimum for ASPCAP).  Our stringent selection criteria, including Gaia DR2 astrometry for later observations, helped maximize the observation of actual members. 

APOGEE-2S has 300 fibers per plate. 
Typically 250 fibers are placed
on science targets, with the remaining 50 divided between 15 standard stars and 35 fibers allocated
to sky. So apart from cluster targets, we are able to observe hundreds of field stars per pointing.
We have no problem filling the remaining non-GC fibers with good targets. Results from these subsidiary projects will be presented in forthcoming papers.

As a Contributed CNTAC APOGEE data set, the CAPOS reductions and pipeline analysis were carried out in the standard way for APOGEE data. Details are presented in \citet{DR16}. In particular, the data are first processed by the APOGEE pipeline \citep{Nidever2015}, which yields heliocentric radial velocity. Then 
stellar parameters are determined using the ASPCAP pipeline \citep{GarciaPerez2018} to derive abundances for some 20 chemical elements for stars with $S/N>70$. 
Line lists are from 
\citet{Shetrone2015} and \citet{Smith2021}.
Each spectrum is analyzed independently in an automated manner.
A detailed analysis of the accuracy and precision of these parameters and abundances is given in \citet{Holtzman2018} and \citet{Henrik2018}. Several improvements have been implemented for DR16, as delineated in \citet{DR16} and further investigated in \citet{Henrik2020}. 
ASPCAP is known to yield more precise abundances for stars with metallicities [Fe/H]$>-1.7$ \citep{Leung2019}. 
All of the BGCs reported on here have H10 metallicities that well exceed this minimum. 

\section{ASPCAP results}

Results
presented here are based on the DR16 ASPCAP analysis. Future papers will explore different analysis techniques such as deriving our own atmospheric parameters and using BACCHUS \citep[Brussels Automatic Code for Characterizing High accUracy Spectra -][]{Masseron2016}, which will allow us to compare  CAPOS results in closer detail with those derived for BGCs observed by the SDSS APOGEE survey and analyzed using this code and independent atmospheric parameters \citep{Masseron2019, Meszaros2020}.  

\subsection{Cluster membership selection}

APOGEE DR16 provides accurate radial velocities (RVs, accurate to typically $\sim$0.05 km $\rm s^{-1}$) and metallicites (accurate within $\sim$0.10 dex) determined by ASPCAP.
In addition, it includes {\it Gaia} DR2 astrometric data, which is available for essentially all our observed stars, although not available for target selection prior to the 2018 observations presented here.
With accurate RVs and metallicities from APOGEE and PMs from {\it Gaia} in hand, CAPOS leveraged both surveys to optimally select bona fide cluster member stars from the original targets observed. 

\begin{figure*}
\centering

   \includegraphics[width=12cm]{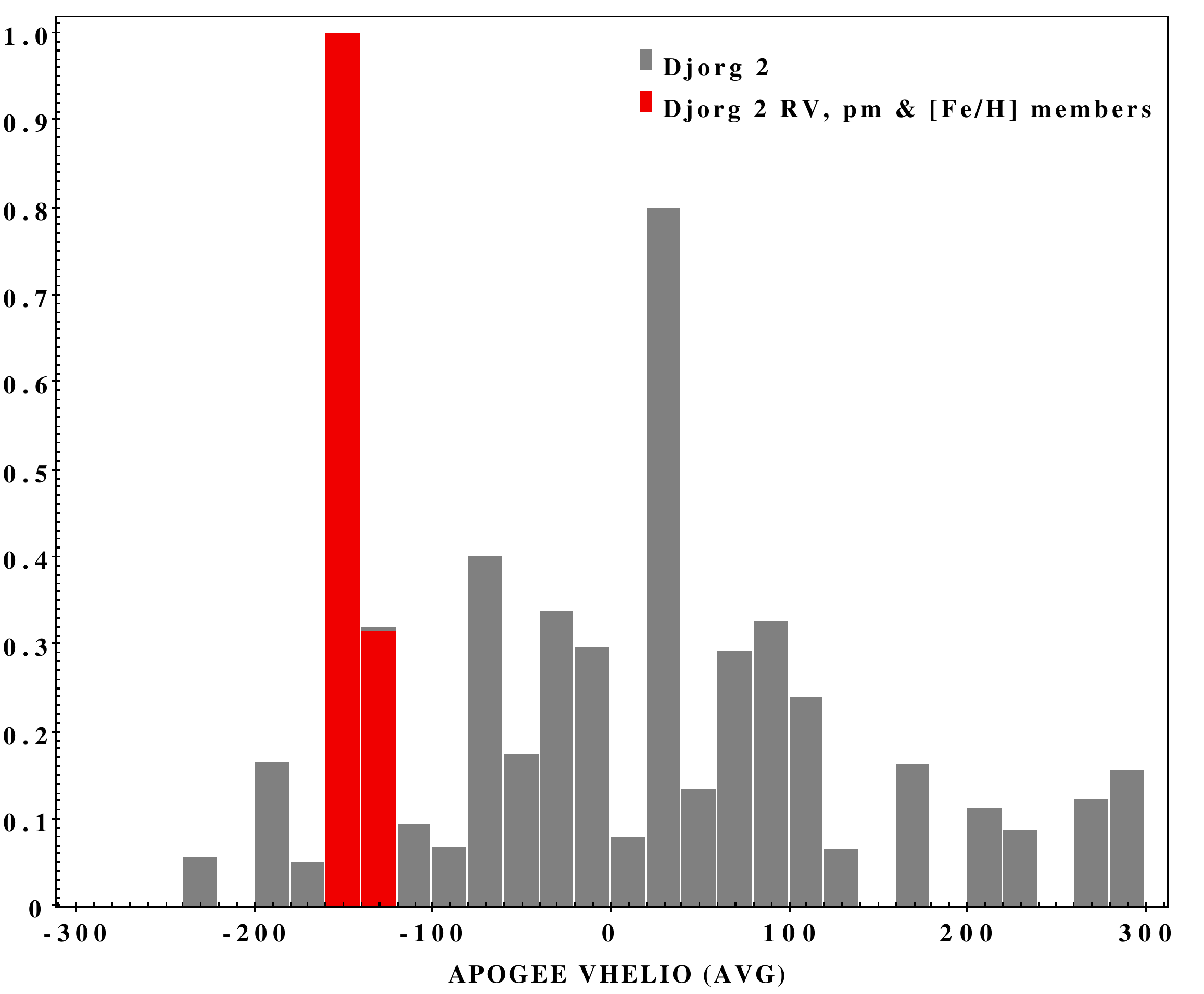}
      \caption{Histogram of ASPCAP heliocentric radial velocity for all stars in the area around the BGC Djorg 2. Our final Djorg 2 members are shown in red, while field stars that did not meet the radial velocity, proper motion, and/or metallicity selection are in gray. }
    \label{membership1}
\end{figure*}

Each CAPOS field reported on here encompassed 2$-$4 cataloged plus candidate clusters in the designated FOV (see Fig. 1).
Targeted stars, including all bulge, K2, EMBLA, and PIGS filler targets observed by APOGEE, were extracted from the DR16 data set by their designated APOGEE field name, then separated into individual cluster target lists.  
Cluster targets with S/N$<$70 were discarded, following previous similar studies \citep[e.g.,][]{Meszaros2020}.
Histograms of APOGEE RVs were constructed for each field, weighted by error, binned and normalized to show the relative number of stars per RV peak per field. An example is shown in 
Fig.\ref{membership1}. This in fact is not a typical case - all other GCs showed single peaks.
It should be noted that many CAPOS clusters have not been extensively studied, and some not at all, so the H10 catalog was consulted only as a guide. Target stars were then isolated within a 3$\sigma$ clip from the mean RV values found in the histograms. In the case of Djorg 2, which shows two peaks (Figure \ref{membership1}), we treated both as possible cluster means until only one met our robust final criteria.

\begin{figure}
   \sidecaption
   \includegraphics[width=10cm]{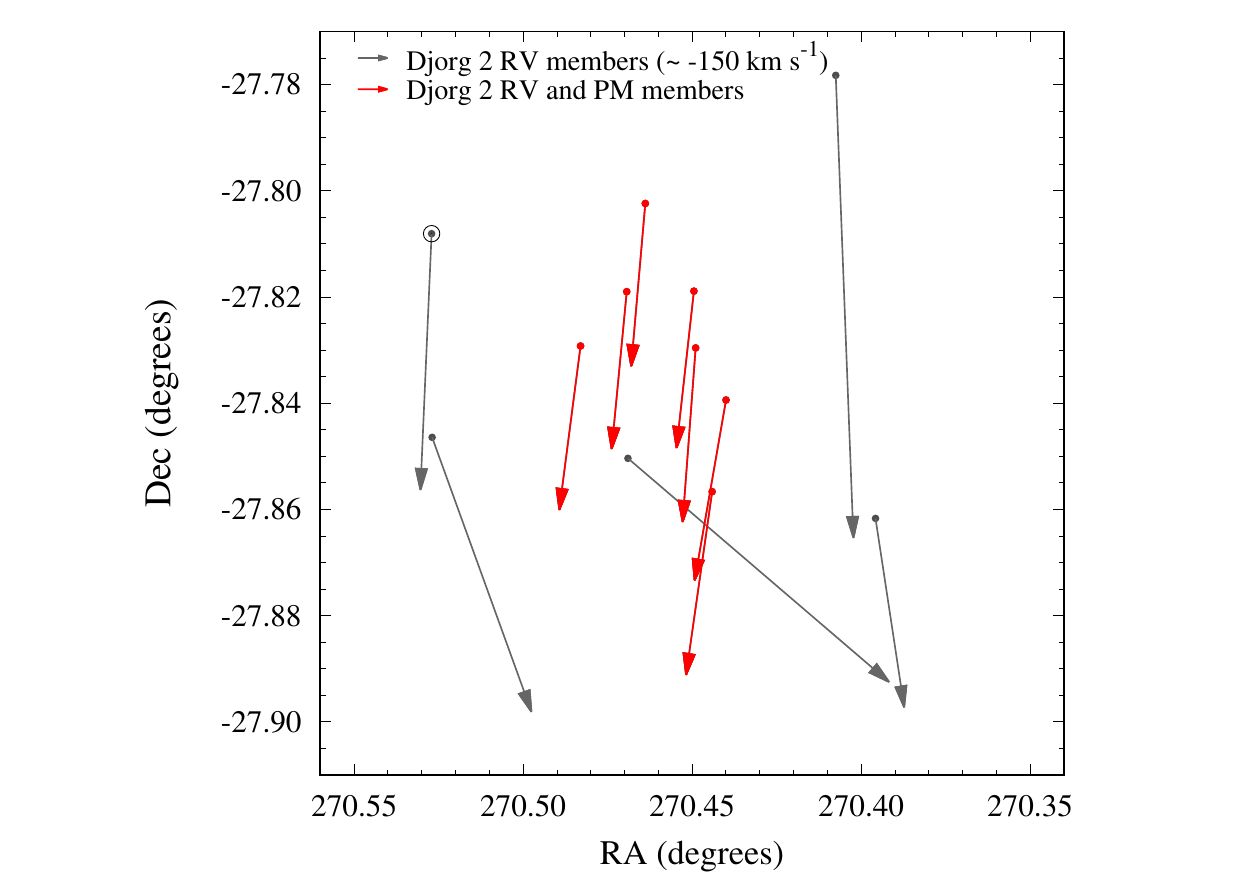}
      \caption{Proper motion vectors on the sky for radial velocity members of Djorg 2. Red corresponds to stars selected as final cluster members from their common radial velocity, proper motion, and metallicity, while gray shows field stars. The black circle indicates a radial velocity and proper motion member that was discarded due to its discrepant metallicity.}
    \label{membership2}
\end{figure}

As a guide to identify each cluster\rq{}s location, density maps in {\it Gaia} PM space were created using the cluster\rq{}s central coordinates in a radius somewhat larger than the half light radius listed in H10. 
The potential RV cluster members were over-plotted onto the same PM space. A second subset of possible members was then isolated in the region identified as the cluster\rq{}s position in the PM density maps. The RV and PM candidate members were then cross matched to create yet another subset of cluster member candidates satisfying both criteria.
To help ensure our candidate member stars are {\it bona fide} cluster members, we include an additional step using {\it Gaia} DR2 PMs by plotting their space motion projected onto the great circle coordinate system. 
Candidates that showed skewed directions from that of the bulk of other candidates were discarded, regardless of RV. An example is shown in Fig.\ref{membership2}.

Finally, we use cluster metallicity as a membership criterion. Given the $\sim$0.10 dex absolute uncertainty for ASPCAP metallicities,
we consider any remaining member candidates as metallicity outliers if their ASPCAP [Fe/H] falls  $>3\sigma$ from
the cluster mean (see Table \ref{meanvalues}). This final criterion resulted in very few candidate members being
discarded, typically one per cluster.

Our method appears to be robust, though restrictive, and yields a high probability our final selections are indeed bona fide cluster members. 
Table \ref{meanvalues} lists our final derived mean cluster parameters and associated standard deviations,
including APOGEE ASPCAP metallicity, [$\alpha$/Fe], and heliocentric RV, our derived mean cluster PM values based on Gaia, {the number of allocated fibers},
the number of final cluster members, and the number of 1G stars (see below). For comparison, Table 2 lists the mean PMs
for CAPOS clusters from \citet{Baumgardt2019} who, unfortunately, do not publish individual cluster member values.
We find the mean cluster PMs between the two studies are in good agreement. We note that the percentage of allocated fibers on a cluster that turned out to correspond to cluster members ranged from 4\% in Terzan 4 to 23\% in HP1, with a average of 12\%.

\begin{table*}
	\begin{center}
		\setlength{\tabcolsep}{2.mm}  
		\caption{CAPOS mean cluster metallicity, [$\alpha$/Fe], radial velocity and proper motion for members.}
		\begin{tabular}{llcrrrrcc}
			\hline
			Cluster ID & [Fe/H]$^a$ & [$\alpha$/Fe]$^{a}$ & $V_{r}$$^{a}$ & $\mu _{\alpha} \cos\delta$$^{b}$& $\mu _{\delta}$$^{b}$& N$_{\rm {fibers}}$ & N$_{\rm members}$ & N$_{1G}$  \\
& (dex) & (dex) & (km $\rm s^{-1}$) & (mas $yr^{-1}$) & (mas $yr^{-1}$) & &  &   \\
			\hline
  Terzan~2 &  $-$0.85$\pm${ 0.04} & 0.26$\pm$0.01 & 133.2$\pm$1.4 & $-$2.23$\pm$0.21 &$-$6.36$\pm$0.16  & { 58} & 4 & 2 \\
Terzan~4 & $-$1.40$\pm${ 0.05} & 0.21$\pm$0.01 & $-$48.3$\pm$3.5 & $-$5.08$\pm$0.26 & $-$2.96$\pm$0.20 & { 74} & 3 & 2 \\
HP~1 &  $-$1.20$\pm${ 0.10} & 0.22$\pm$0.00 & 39.8$\pm$4.0 & 2.57$\pm$0.13 & $-$10.15$\pm$0.10 & { 43} & 10 & 2 \\
Terzan~9 & $-$1.40$\pm${ 0.07} & 0.16$\pm$0.05 & 68.1$\pm$4.3 & $-$2.16$\pm$0.23 & $-$7.39$\pm$0.19 & { 55} & 9 & 4 \\
Djorg~2 &  $-$1.07$\pm${ 0.09} & 0.27$\pm$0.04 & $-$150.2$\pm$4.7 & 0.59$\pm$0.15 & $-$3.17$\pm$0.14 & { 73} & 7 & 4 \\
NGC~6540 & $-$1.06$\pm${ 0.06} & 0.26 & $-$14.4$\pm$1.1 & $-$3.89$\pm$0.14 & $-$2.79$\pm$0.12& { 57} & 4 & 1 \\
NGC~6642 &  $-$1.11$\pm${ 0.04} & 0.25 & $-$56.1$\pm$1.1 & $-$0.17$\pm$0.08 &$-$3.98$\pm$0.06 & { 51} & 3 & 1 \\			
\hline
		\end{tabular}  \label{meanvalues}\\
	\end{center}
	\raggedright{
		$^{a}$ Mean APOGEE [Fe/H], [$\alpha$/Fe], radial velocities, and associated standard deviations for cluster members.\\
		$^{b}$ Mean {\it Gaia} DR2 proper motions and associated errors for CAPOS cluster members.\\
		{\bf Note:} Only cluster members with spectra with S/N $>$70 are included to determine mean values.}
\end{table*}

\subsection{ASPCAP atmospheric parameters}

As part of the pipeline, ASPCAP derives atmospheric parameters simultaneously from a global fit to the entire spectrum, and then detailed abundances for some 20 elements by fitting the spectral lines to models using these atmospheric parameters.
We tested the reliability of the ASPCAP stellar parameters and abundances, first checking for any trend of iron abundance in each cluster as a function of the effective temperature. 
We indeed find a significant positive gradient of increasing metallicity with temperature of similar magnitude for most clusters. 

\begin{figure*}
\centering
   \includegraphics[width=17cm]{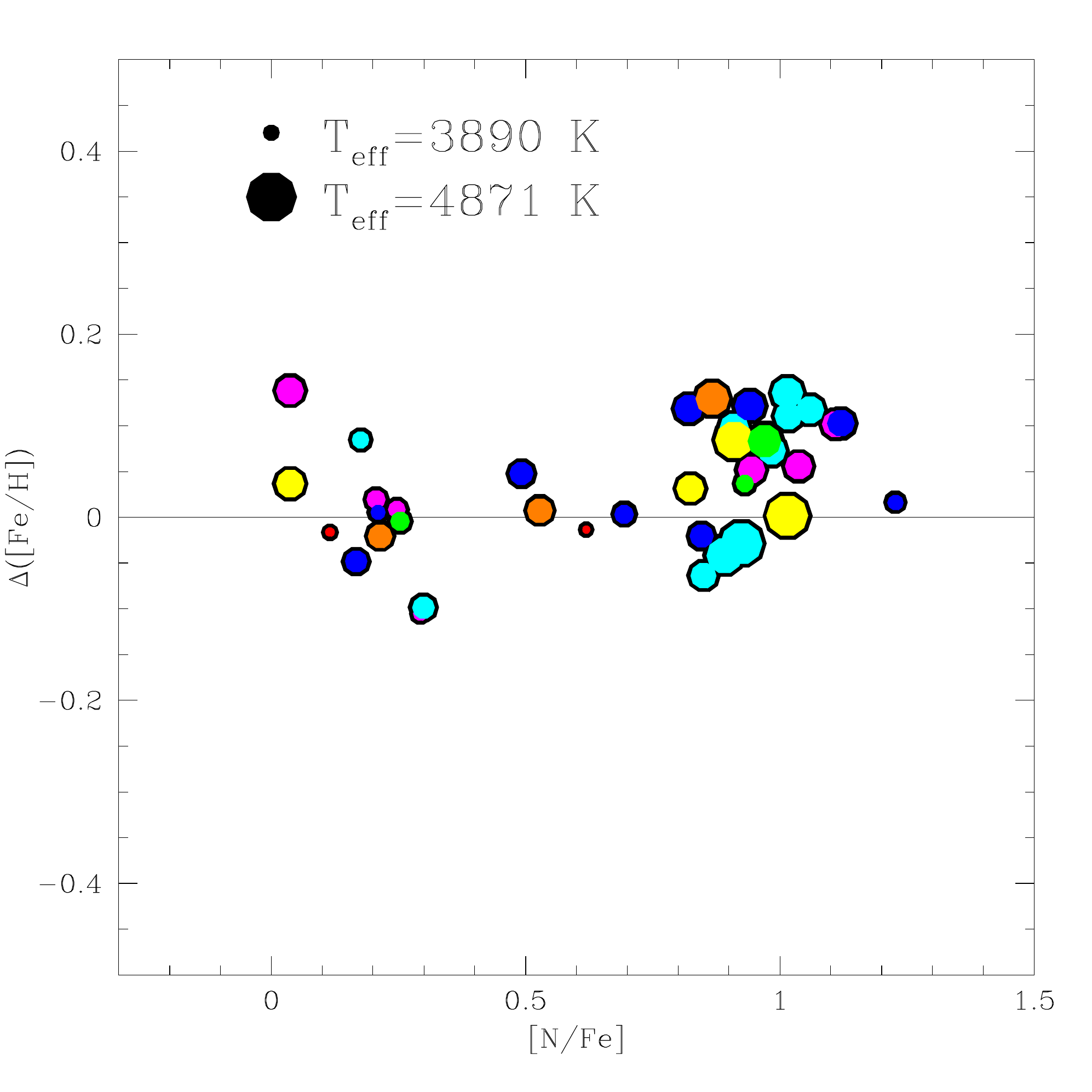}
      \caption{Difference in [Fe/H] as a function of [N/Fe], with the cluster mean for 1G stars ([N/Fe]$<$0.7) taken as the zero point: red = Terzan 2, magenta = Djorg 2, cyan = HP1, yellow = NGC 6540, blue = Terzan 9, orange = Terzan 4, green = NGC 6642.  Symbol size is directly proportional to $T_{\rm eff}$. A clear trend is found for 2G (higher [N/Fe]) stars to have both a higher $T_{\rm eff}$ and [Fe/H] than 1G stars in the same cluster.}
    \label{FevsN}
\end{figure*}

A possible systematic effect on chemical abundances with effective temperature for previous ASPCAP data releases has been studied in the literature, both in GCs \citep{Meszaros2013, Masseron2019}, and field stars \citep[e.g.,][]{Zasowski2019}. In particular,
it is known \citep{Henrik2018} that ASPCAP overestimates (with respect to the optical studies taken as reference) effective temperatures as well as surface gravities for so-called second generation (2G) stars, i.e. stars with abundance patterns typical of stars believed to have been born from gas polluted by  first generation (1G) stars to form a subsequent generation(s), leaving a present-day cluster displaying MP. 

One of the best indicators of 1G versus 2G stars is their N abundance, as N is very substantially
increased from 1G to 2G stars, by an amount that can approach or exceed 1 dex (see below).
Of course, N is also increased by various dredge-up episodes ocurring on the RGB
as a result of normal stellar evolution \citep{Kraft1994, Gratton2000}. However, this does not affect the overall metallicity. 
Fig.~\ref{FevsN} shows that there is indeed a trend within each of our clusters for stars with the highest [N/Fe] abundances (2G stars - hereby defined as those with [N/Fe]$>$0.7) to also show a higher metallicity than their 1G counterparts (and that 2G stars generally have higher $T_{\rm eff}$ than their 1G counterparts). If ASPCAP overestimates $T_{\rm eff}$ for 2G stars, it will also overestimate [Fe/H] since higher $T_{\rm eff}$ generally means fainter spectral lines and a higher metallicity is required to fit a given line. Gravity does not have as great an impact on abundances as $T_{\rm eff}$, at least for giants. It
is expected, given the ASPCAP methodology, that there will be systematic differences for other, possibly all, elements for 2G stars.

For this reason, abundances for 2G stars are typically derived using independent atmospheric parameters,
along with boutique software, such as BACCHUS,
instead of relying on ASPCAP. We will use such a technique in a subsequent paper, but for the purposes of this study we decided to correct the metallicity of 2G stars as follows: We first derive the mean cluster [Fe/H] of all 1G stars. We then calculated the offset of each 2G star from its cluster mean and found a very consistent mean difference of +0.06$\pm$0.01 dex for all 2G stars. We then subtracted this value from the ASPCAP [Fe/H] of all 2G stars, and use this corrected [Fe/H] value to derive the mean metallicity {and its error, which is
the standard deviation convolved with an assessment of the error in this correction}, which are given in Table \ref{meanvalues}. We recognize that, although most stars are either low or high N (1G or 2G, respectively - see Figures 8 and 9), there are some intermediate stars and this is a simplification.  

It is unfortunately very difficult to estimate the effects of this systematic error on the abundances of the other species for 2G stars because the intrinsic dispersion of light elements and the higher measurement errors of heavy elements will blur any underlying trend. 
For this initial paper, we will only use 2G stars in deriving the mean [Fe/H] (in combination with 1G stars) as described here,  and to qualitatively investigate MP, and use only 1G stars to derive mean abundances for all other elements, as given in Tables \ref{meanvalues} ([$\alpha$/Fe]) and \ref{Element}. The number of 1G stars is given in the last column of Table \ref{meanvalues}.

\subsection{ASPCAP abundances}
A number of studies have investigated how well different atomic species are measured with APOGEE spectra, including comparisons of ASPCAP abundances with those from applications of boutique programs to APOGEE spectra as well as to other studies, e.g. from optical spectra. Most recently, \citet{Henrik2020} carry out this procedure for all stars in DR16, while \citet{Masseron2019} and \citet{Meszaros2020} restrict their studies to GCs in the north and south, respectively. The general consensus of these studies is that ASPCAP abundances for 1G stars with metallicity around -1 dex, as is the case for our current sample, are deemed to have well-determined values for at least the following elements: C, N, O, Na, Mg, Al, Si, K, Ca, Fe, Mn, and Ni, so we restrict our current study to these species.

C, N, O, Na, and Al are the key light elements to investigate MPs.
Na is more problematic than the other elements listed in terms of precise measurement with ASPCAP. There are only two Na I lines, which are very weak, especially at the relatively low metallicities of our sample (although they are seen in the coolest stars in M107, with a metallicity of $-$1: \citet{Masseron2019}, and located in a region heavily blended by telluric features. Nevertheless, we keep Na due to its important role in investigating MP, but emphasize that the results should be viewed with caution. \citet{Fernandez-Trincado2020} investigate differences between ASPCAP and BACCHUS abundances for some 1000 presumed 1G field stars with -2 $<[Fe/H]<-0.65$. They found ASPCAP to overestimate the abundances of N, O, and Al by about 0.2 dex with a scatter of about 0.1 dex. Nevertheless, we will use the ASPCAP abundances at face value.

As for the $\alpha$-elements, which  traditionally include O, Mg, Si, S, Ca, and Ti, all are measured with ASPCAP. In addition, ASPCAP provides an overall estimate of a global ($\alpha$-elements relative to metal abundance), denoted [$\alpha$/M], essentially [$\alpha$/Fe], in its fit to determine the atmospheric parameters, fitting all of the $\alpha$-elements simultaneously  while keeping their relative abundances identical. We consider O as an element strongly affected by MP \citep[and indeed ASPCAP  O abundances are particularly problematic for 2G stars -][]{Masseron2019}.  Mg and Si can also be affected by MP, but to a lesser extent \citep[e.g.,][]{Bastian2018}. We further note that both S and Ti are not well measured by APOGEE \citep{Henrik2020}. So we are left with Mg, Si, and Ca as the best APOGEE $\alpha$ tracers. Happily, all three of these key $\alpha$-elements are very precisely measured in APOGEE. \citet{Nidever2020} investigate how ASPCAP abundances of these elements compare to the high-resolution optical studies of \citet{Carretta2009} for 58 1G giants in 11 GCs with a wide range of metallicity, covering the range of our BGCs studied here. They found a mean offset of 0.17 dex for [Mg/Fe]
with a standard deviation of 0.08 dex, an offset of 0.09 dex for [Si/Fe] with a standard deviation of 0.07 dex, and an offset of 0.16 dex for [Ca/Fe] with a standard deviation of 0.16 dex, with all of the offsets being in the same sense - with the ASPCAP values lower than the Caretta values. \citet{Fernandez-Trincado2020} investigate Mg and Si abundances in their large field star sample and find small mean ASPCAP overestimates but null within the errors. Again, we will use the ASPCAP abundances at face value. 
 
\citet{Nidever2020} also verified that the parameter-level [$\alpha$/M] value yields abundance patterns consistent with those of the individual $\alpha$ elements but is more precise, and therefore seems the current best choice for the ASPCAP [$\alpha$/Fe] abundance. 
There is currently some freedom to choose the best estimate of the overall [$\alpha$/Fe] abundance in APOGEE, including the parameter-level [$\alpha$/Fe]
\citep[e.g.,][]{Nidever2020}, [Mg/Fe] \citep[e.g.,][]{Rojas-Arriagada2019}, [Si/Fe] \citep[e.g.,][]{Horta2020}, or some combination thereof. 

The fundamental element Fe is well measured in ASPCAP. We note that ASPCAP metallicities, at least for 1G stars in other GCs, are generally in good agreement with those of other high-resolution studies. \citet{Nidever2020} compare the metallicities of stars in 26 GCs with ASPCAP metallicities ranging from -0.6 to -2.3 with those of other high-resolution studies, and found a mean offset of 0.06 dex to higher metallicity for ASPCAP and a scatter of 0.09 dex, while \citet{Fernandez-Trincado2020} find an offset of 0.11$\pm$0.11 dex in the opposite sense when comparing ASPCAP to BACCHUS abundances. We will simply use the ASPCAP values, after correcting the metallicities of 2G stars as described above. Our mean [Fe/H] values are given in Table \ref{meanvalues}. 
Of the various Fe-peak elements measured by ASPCAP, including 
V, Cr, Mn, Co, Ni, and Cu, the most reliable are Mn and Ni \citet{Henrik2020}, so we include only these here. 

We include K in our analysis but only give our mean values, saving details for this and other elements for a future paper. 
The mean abundances and their standard deviations from 1G stars are reported in Table \ref{Element} for the 11 elements we study besides Fe and $\alpha$. We note that we exclude from cluster means the very rare cases where an element in a given star was flagged as being unreliable. 

\subsection{Fundamental cluster parameters: Mean metallicities, [$\alpha$/Fe] abundances, and radial velocities}
The mean metallicity, designated [Fe/H], of a GC is the primary parameter detailing its chemical composition. The next most salient composition indicator is the abundance of the  $\alpha$ elements, designated as [$\alpha$/Fe].
Finally, RVs provide crucial information regarding membership, internal kinematics and the cluster's orbit. Despite the critical
importance of these parameters, among the most fundamental for our understanding of a cluster's formation and subsequent chemical and dynamical evolution, the current state of our knowledge of these parameters for BGCs is woefully inadequate. This is particularly true for most of our current sample. 

\begin{table*}
	\begin{center}
		\setlength{\tabcolsep}{2.mm}  
		\caption{Mean abundances of first generation stars}
		\begin{tabular}{ l c  c c  c  c c  c   }
			\hline
			\hline
Element   & Terzan 2      &  Terzan 4       &  HP 1          &    Terzan 9       &  Djorg 2        &   NGC 6540      &   NGC 6642     \\
\hline
${\rm [C/Fe]}$   & -0.03      	  &  -0.30     	    &  -0.24         &    -0.37	  	 &  -0.11	   &   -0.06         &   -0.13        \\
& $\pm$0.03  	  &  $\pm$0.01 	    &  $\pm$0.05     &    $\pm$0.05  	 &  $\pm$0.10	   &   $-$     &   $-$    \\
${\rm [N/Fe]}$    & 0.37	  &  0.37           &  0.24          &    0.39       	 &  0.20	   &   0.04          &   0.25         \\
& $\pm$0.25  	  &  $\pm$0.16 	    &  $\pm$0.06     &    $\pm$0.21  	 &  $\pm$0.10	   &   -     &   -    \\
${\rm [O/Fe]}$    & 0.28       	  &  0.24      	    &  0.23          &    0.19       	 &  0.30       	   &   0.33          &   0.28         \\
& $\pm$0.00  	  &  $\pm$0.01 	    &  $\pm$0.00     &    $\pm$0.04  	 &  $\pm$0.04	   &   -     &   -    \\
${\rm [Na/Fe]}$   & 0.01	  &  0.15           &  -0.17         &   -0.07	  	 &  -0.19	   &   0.28          &   0.18	      \\ 
& $\pm$0.03  	  &  -	    &  $\pm$0.02     &    $\pm$0.20  	 &  $\pm$0.15	   &   -     &   -    \\
${\rm [Mg/Fe]}$   & 0.38	  &  0.25           &  0.27          &    0.22	  	 &  0.34       	   &   0.29          &   0.34	      \\ 
& $\pm$0.05  	  &  $\pm$0.00 	    &  $\pm$0.01     &    $\pm$0.06  	 &  $\pm$0.04	   &   -     &   -    \\
${\rm [Al/Fe]}$   & 0.05       	  &  -0.14      	    &  -0.14          &    -0.09       	 &  0.02	   &   -0.03          &   0.10	      \\ 
& $\pm$0.01  	  &  $\pm$0.12 	    &  $\pm$0.03     &    $\pm$0.22  	 &  $\pm$0.11	   &   -     &   -    \\
${\rm [Si/Fe]}$   & 0.25 &  0.23           &  0.27          &    0.21       	 &  0.29	   &   0.27          &   0.35         \\
& $\pm$0.01  	  &  $\pm$0.02 	    &  $\pm$0.01     &    $\pm$0.04  	 &  $\pm$0.04	   &   -     &   -    \\
${\rm [K/Fe]}$    & -       	  &  0.17      	    &  0.12          &    0.30       	 &  0.20       	   &   0.23          &   0.20         \\
& -  	  &  $\pm$0.01 	    &  $\pm$0.15     &    $\pm$0.13  	 &  $\pm$0.02  	   &   -     &   -    \\
${\rm [Ca/Fe]}$   & 0.26       	  &  (0.02)      	    &  0.22          &    0.22       	 &  0.19       	   &   0.16          &   0.24         \\
& $\pm$0.01  	  &  $\pm$0.17 	    &  $\pm$0.05     &    $\pm$0.05  	 &  $\pm$0.06  	   &   -     &   -    \\	  
${\rm [Mn/Fe]}$   & -      	  &  -0.26     	    &  -0.29         &    -0.19      	 &  -0.27      	   &   -0.35         &   -0.34        \\
& -  	  &  $\pm$0.04 	    &  $\pm$0.05     &    $\pm$0.05  	 &  $\pm$0.03  	   &   -     &   -    \\
${\rm [Ni/Fe]}$   & 0.02       	  &  -0.01     	    &  -0.07         &    -0.01      	 &  0.00       	   &   -0.03          &   0.05         \\
& $\pm$0.00  	  &  $\pm$0.08 	    &  $\pm$0.05     &    $\pm$0.03  	 &  $\pm$0.03  	   &   -     &  -    \\
\hline
		\end{tabular}  \label{Element}\\
	\end{center}
	\raggedright{}
\end{table*}

Indeed, CAPOS was devised to address this problem for as many BGCs as possible, taking advantage of the powerful APOGEE instrument, which was designed to deliver high-precision RVs and abundances for a large number of elements, including Fe and all of the species considered $\alpha$-elements. In this section we discuss the ASPCAP results for [Fe/H], [$\alpha$/Fe], and RV for our initial CAPOS clusters. 

\subsubsection{Metallicity}
We first discuss Fe. Fe is generally synonymous with metallicity.
The metallicity of a cluster has long been recognized as an excellent tracer of its nature, and gives invaluable insight into the cluster's supernova enrichment history. \citet{Zinn1985} first divided Galactic GCs into halo and disk systems primarily based on [Fe/H]. Then \citet{Minniti1995a} argued that some of the inner metal-rich GCs belong to the GB. More recently, \citet{Bica2016} used [Fe/H] to discriminate between BGCs and non-BGCs, with a division at -1.5. They noted that the metallicity distribution for BGCs show two peaks - the traditional one associated with BGCs at [Fe/H]=-0.5 but also another one around -1, which in fact is of equal if not greater strength. They point out that these lower metallicity clusters lie at the low end of the bulge field-star metallicity distribution and are the best candidates for the oldest Galactic objects. It turns out that all of our present CAPOS sample are members of the lower-metallicity subset. Unfortunately, the metallicity information for our clusters, and  indeed most BGCs, 
comes from a hodgepodge of sources with a large range of precision and accuracy, but mostly of relatively poor reliability and not involving near-IR capabilities to overcome the extinction problem.
Although all of our clusters have been investigated before, at least in terms of metallicity and radial velocity,  the metallicity estimates are generally based on relatively low-quality indices, such as the slope or color of the RGB in a CMD or low-resolution optical spectra. Only two of the sample have been the subject of high-resolution spectroscopic studies, and only one of these within the last 15 years. Needless to say, such studies are very inhomogeneous, and make it very difficult to compare abundances in one cluster derived with one method to that in another derived with a different method. 
CAPOS now provides unprecedented metallicities, of much higher quality than virtually all previous estimates and on a homogeneous scale, allowing a quantum leap forward in our knowledge of these fascinating, but until now very poorly studied objects, and of the BGC system in general.

We address each of our clusters in turn. First, we note that
none of them show any evidence for internal metallicity variations, although the sample size in each cluster is relatively small, varying from 3-10 members. The errors quoted in Table 2 are standard deviations, and are all consistent with expected observational errors.
The bible of Galactic GC properties, including metallicity, is H10, so we begin with his references and also include more recent studies. Finally, we note that \citet{Horta2020} investigated several of our clusters using the same DR 16 ASPCAP data, and obtained virtually identical results, the only difference resulting from their slightly different membership for one cluster and the fact that they used uncorrected metallicites for any 2G stars, so that their metallicities are generally slightly higher than ours. Their emphasis was on distinguishing in situ versus accreted GCs chemically using APOGEE data, and included a total of 46 GCs, and did not focus on the details of the chemistry for any cluster. We quote their values where appropriate, but emphasize that they are not independent from ours. \\

{\bf Terzan 2:} 
The H10 [Fe/H] value for this cluster is -0.69, based on low-resolution Ca triplet (CaT) integrated spectroscopy \citep{Armandroff1988} and near-IR low-resolution spectra of seven stars \citep{Stephens2004},
while our mean value is -0.85$\pm$0.02 from four stars \citep[][derive a slightly higher value of -0.82]{Horta2020}. In the literature, one finds values ranging from -0.25 \citep{Kuchinski1995} to -1.07 \citep{Stephens2004}, a range of over 0.8 dex. No other high-resolution abundance study has been carried out. Among the most reliable, i.e. recent spectroscopic studies, are those of  \citet{Dias2016}, who found -0.72$\pm$0.13 from low-resolution optical spectra, and several studies using the CaT technique, including \citet{Vasquez2018}, who yield values ranging from -0.42 to -0.68 depending on their calibration, and \citet{Geisler2021}, who give -0.65$\pm$0.03.
Our value is in reasonable agreement with H10 and 
in generally good to reasonable agreement with low-resolution spectroscopic studies, although more metal-poor. 

{\bf Terzan 4:}
The H10 [Fe/H] value for this cluster is -1.41, based on low-resolution CaT integrated spectroscopy \citep{Armandroff1988},  near-IR low-resolution spectra of seven stars \citep{Stephens2004}, and high-resolution spectra of four stars \citep{Origlia2004}, while our mean value is -1.40$\pm$0.04 from three stars. In the literature, one finds values ranging from -1.41 (H10) to -1.60, a comfortably small range. This latter value comes from \citet{Origlia2004}, which is the most reliable previous study as it is based on high-resolution near-IR spectra of four stars using the Keck NIRSPEC instrument. They also derive abundances for several other elements, which we discuss in detail below.
Our value is virtually identical to that of H10, and in good agreement
with the other high-resolution spectroscopic determination, although more metal-rich.

{\bf HP1:}
The H10 [Fe/H] value for this cluster is -1.00, based on low-resolution CaT integrated spectroscopy \citep{Armandroff1988},  near-IR low-resolution spectra of two stars \citep{Stephens2004} and high-resolution spectra of two stars \citep{Barbuy2006}, while our mean value is -1.20$\pm$0.08 from ten stars \citep[][give -1.14 from 12 stars]{Horta2020}. In the literature, one  finds values ranging from -0.56 \citep{Minniti1995b} to -1.60 \citep{Davidge2000}, a range of over 1 dex. Two high-resolution VLT UVES
studies have been carried out by \citet{Barbuy2006, Barbuy2016}. In the latter paper, they combine their sample for a total of 8 stars, deriving a mean metallicity of -1.06$\pm$0.15.
They also derive abundances for several other elements, which we discuss in detail below.
Our value is in good agreement with their value and in comfortable accord with that of H10, although more metal-poor.

{\bf Terzan 9:}
The H10 [Fe/H] value for this cluster is -1.05, but relatively poorly determined, with a low weight, based only on near-IR photometry \citep{Valenti2010}, while our mean value is -1.40$\pm$0.05 from nine stars. In the literature, one finds values ranging from -0.38 \citep{Zinn1985} to -2.1 \citep{Ortolani1999}, an enormous range of 1.7 dex. No other high-resolution abundance study has been carried out. Among the most reliable studies are those of  \citet{Dias2016}, who found -1.06$\pm$0.13 from low-resolution optical spectra, the \citet{Ernandes2019} value of -1.10$\pm$0.15, also from low-resolution optical as well as near-IR spectra, and several studies using the CaT technique,
including \cite{Vasquez2018}, who yield values ranging from -1.08 to -1.21 depending on their calibration, and 
\citet{Geisler2021}, who give -1.12$\pm$0.03.
Our value is in relatively poor  agreement with both H10 
and most of the low-resolution spectroscopic studies, falling to lower metallicity than all of them.

We note that the optical results from \citet{Ernandes2019} actually appear quite bimodal, with a peak near -1.4 and another near -0.7, with the former agreeing with our value. Given their mean metallicity of -1.1, and the relatively blue HB suggested by the ground-based optical CMD of \citet{Ortolani1999}, they found Terzan 9 to be a good candidate for a very old GC. However, we find a significantly lower metallicity. At the same time, the deep HST near-IR CMD from \citet{Cohen2018} does not reveal a very significant BHB, if any.

{\bf Djorg 2:}
The H10 [Fe/H] value for this cluster is -0.65 but very poorly determined, based only on near-IR photometry \citep{Valenti2010}, while our mean value is -1.07$\pm$0.07 from seven stars. In the literature, one finds values ranging from -0.3 \citep{Bica1998} to -1.11 \citep{Ortolani2019}, a range of 0.8 dex. No other high-resolution abundance study has been carried out. 
However, we note that \cite{Kunder2020} have studied this cluster using the same DR16 ASPCAP database. They identify the same members as we do but derive a slightly higher mean metallicity of -1.05 since they do not correct the 2G stars.
Among the most reliable studies are those of   
\cite{Dias2016}, who found -0.79$\pm$0.09 from low-resolution optical spectra, and several studies using the CaT technique,
including \cite{Vasquez2018}, who yield values ranging from -0.97 to -1.09 depending on their calibration, and 
\citet{Geisler2021}, who give -0.75$\pm$0.05.
Our value is in poor agreement with H10, but in good to fair agreement with the recent low-resolution spectroscopic studies, although generally more metal-poor. 

{\bf NGC 6540:}
The H10 [Fe/H] value for this cluster is -1.35, but relatively poorly determined, based on high-resolution but low S/N optical spectra of six stars \citep{Cote1999}, while our mean value is -1.06$\pm$0.04 from four stars (Horta et al. 2020a derive -1.01 from the same sample). In the literature, one finds values ranging from -1.01 to -1.5 \citep{Vulic2018}, a range of half a dex. No other high-resolution abundance study has been carried out. The only other spectroscopic study, using the CaT technique, is that of \citet{Geisler2021}, who give -1.05$\pm$0.05.
Our value is in reasonable agreement with H10, and in excellent  agreement with the recent low-resolution spectroscopic study.

{\bf NGC 6642:}
The H10 [Fe/H] value for this cluster is -1.26, based on a recalibration of the metallicity derived from low-resolution optical spectra of five stars \citep{Minniti1995b}, while our mean value is -1.11$\pm$0.02 from three stars. In the literature, one finds values ranging from our value to -1.29 \citep{Minniti1995b},
the smallest range among our sample. No other high-resolution abundance study has been carried out. Among the most reliable studies are those of \citet{Minniti1995b} and  the CaT study of \citet{Geisler2021}, who derive -1.15$\pm$0.05.
Our value is in good agreement with all other values, especially the recent low-resolution spectroscopic study. 

The bulge field star metallicity distribution measured by ASPCAP has been recently studied by \citet{Rojas-Arriagada2020}. They compile a total of $\sim$13000 bulge stars and find strong evidence for trimodality, with peaks at [Fe/H] = +0.32, -0.17 and -0.66. These peaks maintain their value but their strength varies as a function of Galactic latitude. The fraction of stars below -1 is very small,
in contradistinction to our sample, all of which fall well below the most metal-poor field peak. Of course, our limited sample is not representative of the whole BGC system. The metallicity distribution function of BGCs has been shown by \citet{Bica2016} to be bimodal, with peaks around -0.5 and -1.1, using H10 values. The metal-rich peak is well-known as most studies refer to BGCs as metal-rich, with a peak around this value, but as Bica et al. demonstrate the metal-poor peak is perhaps dominant. All of our sample belong to the metal-poor distribution. It is likely that the most metal-poor field-star peak and metal-rich GC peak have similar origins. However,
it is unclear why the field and GC metallicity distributions are otherwise quite distinct, but one of the goals of CAPOS is to derive accurate, homogeneous metal abundances for as large a sample of BGCs as possible to help investigate this issue.

\subsubsection{The $\alpha$-elements}
The $\alpha$-elements play a crucial role in divulging the chemical evolution of a system, in particular the past rate of star formation, as well as information on the IMF. In  concert with the metallicity, they reveal the onset of the dominance of SNeIa over SNeII.
In addition, a detailed knowledge of the [$\alpha$/Fe] abundance ratio is critical to derive an accurate age estimate from a GC deep CMD \citep[e.g.,][and references therein]{Catelan2018}. 

As noted above, there is some freedom currently as to which particular element or combination thereof to use to best represent the $\alpha$ abundance from APOGEE data. We note that all four different choices investigated here - global $\alpha$ (Table \ref{meanvalues})
and Mg, Si, and Ca (Table \ref{Element}) - are all very well-determined in all of our sample, with typical errors of the mean of only a few 0.01 dex, with the lone exception of Ca in Terzan 4, which has a large spread and whose mean value is very uncertain. All four mean values for a given GC are also in very good accord, with the different means falling within 0.04 - 0.15 dex (with the above exception). There are some small systematic offsets - on average for our sample, the highest mean cluster abundance is that of [Mg/Fe], followed by [Si/Fe] (0.03 dex lower), [$\alpha$/Fe] (0.07 dex lower),  and [Ca/Fe] (0.09 dex lower, excluding Terzan 4).
These comparisons demonstrate that the various ASPCAP measurements of the $\alpha$-element abundances are similar to within 0.1 dex.

In addition, we find that the range of mean values of the four abundances amongst our sample is also small: The  [$\alpha$/Fe] abundances of our CAPOS clusters all lie within 0.11 dex, [Mg/Fe] within 0.16 dex, [Si/Fe]  within 0.14 dex, and [Ca/Fe] within 0.1 dex (again excluding Terzan 4). Given the thorough investigation by \citet{Nidever2020}, we prefer the more robust measure of the 
parameter-level [$\alpha$/Fe] abundance as the best representation of the $\alpha$ abundance. However, we note that this abundance
produces an unexpected ”finger”  for metal rich, [$\alpha$/Fe]=0.2 stars that is not produced in the corresponding [Mg/Fe] plot \citep{Henrik2020}, and hence Mg is the preferred representative for $\alpha$-elements for very metal-rich stars, although this is irrelevant for our sample.

There is very little in the literature on $\alpha$ abundances for our sample. \citet{Horta2020} in fact derive mean [Si/Fe] values for some of our GCs from the same DR16 data, but do not distinguish between 1G and 2G stars, and also have a slightly different membership list for HP1. Given that Si is also affected to some extent by MPs, and that 2G stars generally have smaller [Si/Fe] than 1G stars (see below), we expect the Horta mean [Si/Fe] values, using all stars irrespective of their 1G or 2G nature, to be somewhat lower than our means, using only 1G stars, and their errors to be larger. For the three GCs in common, Horta gives [Si/Fe] = 0.26$\pm$0.02 for Terzan 2, while we derive 0.25$\pm$0.01; 0.22$\pm$0.06 for HP 1 versus our 0.27$\pm$0.01, and
0.21$\pm$0.04 for NGC 6540 versus our 0.27 (only one 1G star). This generally agrees with our expectations. 
In  addition, from the same DR16 ASPCAP database and for the same sample of stars as ours, \cite{Kunder2020} derive [Si/Fe] = 0.25 for Djorg 2, 0.04 dex lower than our value, again as expected given their inclusion of 2G stars.  
\citet{Ernandes2019} derive [Mg/Fe] = 0.27$\pm$0.03 for Terzan 9 from low-resolution optical spectra, in accord with our value of 0.22$\pm$0.03.
A comparison with the only other independent high-resolution studies, including other elements besides $\alpha$s,  will be described in more detail below.

\begin{figure*}
\centering
   \includegraphics[width=17cm]{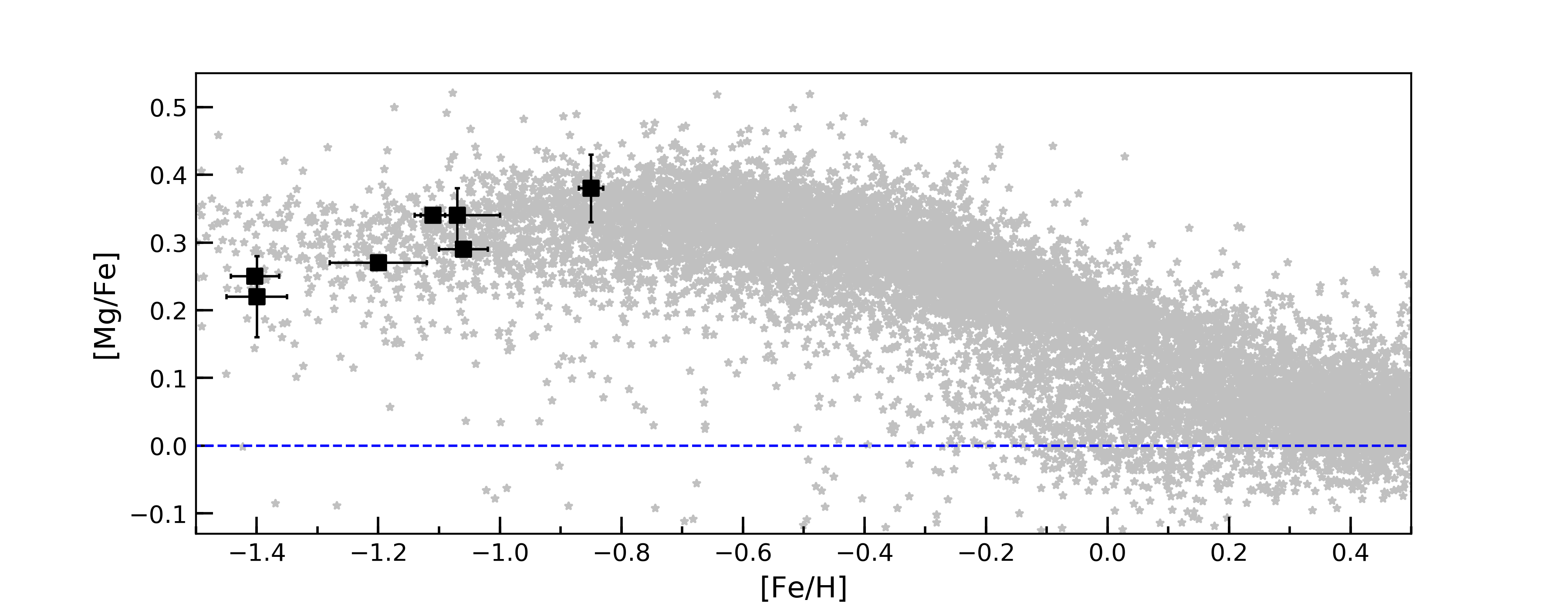}
      \caption{Mean [Mg/Fe] versus [Fe/H] for our CAPOS sample
       (filled squares with error bars), compared with the general trend of APOGEE bulge stars \citep[asterisks - ][]{Rojas-Arriagada2020}.
}
    \label{MgFe}
\end{figure*}

Fig \ref{MgFe} shows how our mean cluster  [Mg/Fe] versus [Fe/H] values compare to APOGEE bulge field stars from \citet{Rojas-Arriagada2020}. All of our BGCs are relatively metal-poor and all fall along the high-$\alpha$ (low-SNeIa) sequence. They also generally follow closely the field star trend at comparable metallicity. However, we note that our two most metal-poor BGCs - Terzan 4 and Terzan 9 - fall somewhat below the mean field-star trend at this low metallicity. Indeed, the high-$\alpha$ field stars form a plateau for metallicities below [Fe/H]$\sim$-0.5, with a value of [Mg/Fe] $\sim$+0.33, which our metal-richest BGCs share. However, our three most metal-poor GCs fall below this value, especially the two most metal-poor. It would be of great interest to see if this trend continues to lower metallicity, although the number of both field stars and clusters should be very small
below -1.4.

Next we follow the recent analysis of \citet{Horta2020}, who use Si as a proxy for the $\alpha$ abundance to investigate all GCs well-measured with APOGEE in DR16. They included several of our clusters in their independent analysis and derived similar values to ours for [Si/Fe], though not identical due to small membership differences and our exclusion of 2G stars. Although we prefer to use the global [$\alpha$/Fe], for consistency sake we here use our [Si/Fe] values. We recall that these are within 0.04 dex of [$\alpha$/Fe] on average. 
Horta compared their APOGEE sample of Main Bulge (MB) and Main Disk (MD) GCs, presumably formed in situ, to GCs associated with various accretion events, including Sequoia (Seq), Helmi streams (H99), and Gaia-Enceladus (GE).  They found that the in situ GCs have [Si/Fe] abundances slightly higher than their presumably accreted counterparts at the same metallicity.

\begin{figure*}
\centering
   \includegraphics[width=17cm]{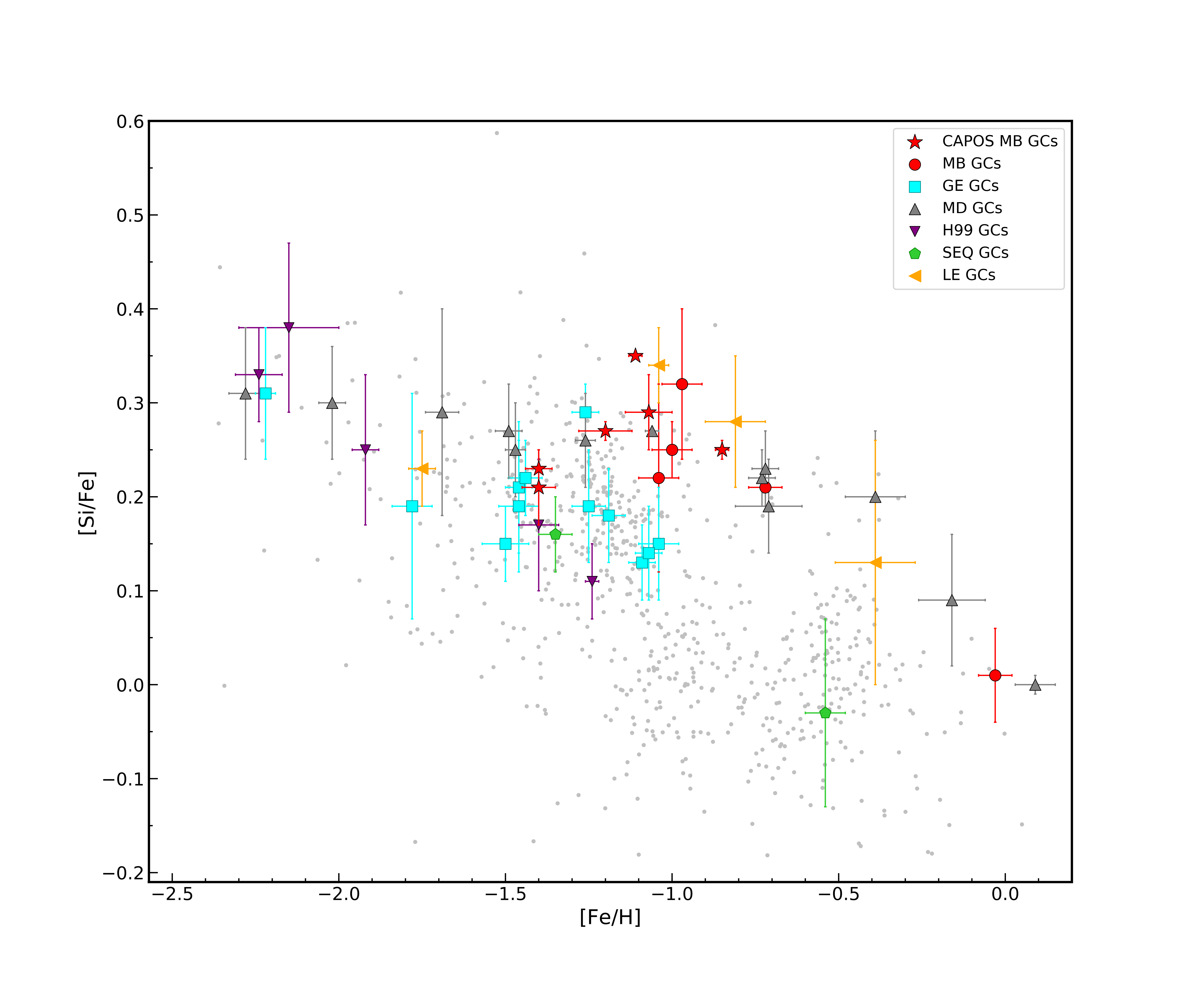}
      \caption{Mean [Si/Fe] versus [Fe/H] for GCs from CAPOS  (red stars), other Main Bulge GCs (red circles),  Gaia-Enceladus (cyan squares),  Main Disk (gray triangles), Helmi streams (purple triangles), Sequoia (green pentagons) and LE GCs (yellow triangles). Values for non-CAPOS clusters are from \citet{Horta2020}. Gray points show the halo field population defined as in \citet{Massari2019}.}
    \label{SiFe}
\end{figure*}

In Fig.\ref{SiFe} we add our CAPOS BGC mean [Si/Fe] values to the data from Horta et al. (2020), where we have used our values for the three clusters in common. We use the updated assessment of GC to a given accretion event from \cite{Bajkova2020}, and also now distinguish Main Bulge from Main Disk GCs. We note that one of our BGCs, NGC 6540, has been modified from a Main Bulge cluster \citep{Massari2019} to a Main Disk cluster.
We use the same field comparison population as \cite{Horta2020}.
We note that CAPOS significantly increases the sample of Main Bulge GCs and also extends them to lower metallicity.
The interesting trend they found is confirmed by using our larger sample.  They derived a mean [Si/Fe] for their sample of six in situ GCs lying between [Fe/H] = -1 to -1.5, where there are a reasonable number of both in situ and accreted cluster types, of 0.25$\pm$0.03 versus 
0.17$\pm$ 0.05 for their 12 accreted GCs in the same metallicity range. We add a number of in situ GCs and also include NGC 6544 as an accreted GE GC \citep{Bajkova2020} and find these values now become 
0.26$\pm$ 0.04 and 0.18$\pm$ 0.04, for 11 in situ and 13 accreted GCs, respectively.
Thus, the significant difference in [Si/Fe] between in situ and accreted GCs persists.
We note that the one Low Energy (LE) GC in this range has a very high value, suggesting an in situ origin.

The general explanation for this difference is that the lower potential well present in the dwarf galaxy progenitors where the accreted GCs formed led to a lower star formation and chemical enrichment rate, and therefore lower [$\alpha$/Fe] than in the deeper potential of the main Galaxy. We note that this trend may not continue below a metallicity of about -2, where the sample size is small and errors are large, and indeed no BGCs exist. However, in field stars we see that the chemical differences in [$\alpha$/Fe] between dwarf galaxy populations and those in the Milky Way largely disappear at low metallicities \citep[e.g.,][]{Tolstoy2009}.
We note that there are growing hints for chemical evolution differences between accreted and in situ populations; e.g., a very recent study finds that GE stars clearly show higher ratios of [Eu/Mg] than in situ stars but
do not show enhanced [Ba/Eu] or [La/Eu] ratios, suggesting an increase in r-process but a lack of significant s-process contribution in the progenitor \citep{Matsuno2021}. APOGEE provides several s-process species we can explore in the future.

\begin{table*}
	\begin{center}
		\setlength{\tabcolsep}{2.mm}  
		\caption{Comparison of abundances with other high-resolution studies}
		\begin{tabular}{l c c c c  }
			\hline
Element    &     HP 1           &    HP 1     &      Terzan 4       &  Terzan 4\\
$ [$X/Fe$] $&   This work        &    B16      &    This work       &   OR04\\
\hline
C          &   -0.24$\pm$0.05   &    0.00     &   -0.30$\pm$0.01   &  -0.25\\
N          &   +0.24$\pm$0.06    &    +0.53     &   +0.37$\pm$0.16   &   -\\
O          &   +0.23$\pm$0.00    &    +0.40     &    +0.24$\pm$0.01   &  +0.54\\
Na         &   -0.17$\pm$0.02   &    -0.16    &    +0.15  &  -\\
Mg         &   +0.27$\pm$0.01    &    +0.36     &    +0.25$\pm$0.00   &  +0.41\\
Al         &   -0.14$\pm$0.03    &    +0.04     &    -0.14$\pm$0.12   &  -\\
Si         &   +0.27$\pm$0.01    &    +0.27     &    +0.23$\pm$0.02   &  +0.55\\
Ca         &   +0.22$\pm$0.05    &    +0.13     &    (+0.02)$\pm$0.17   &  +0.54\\
			\hline
		\end{tabular}  \label{comparison}\\
	\end{center}
	\raggedright{{\bf Note:} \textbf{References.} B16: \citet{Barbuy2016}; 
		OR04: \citet{Origlia2004}.}
\end{table*}

\subsubsection{Comparison with other high-resolution studies}
	  
We next compare our abundances with those of other high-resolution spectroscopic studies. Only two of our clusters have been studied at high resolution,
HP1 and Terzan 4. \citet{Barbuy2006, Barbuy2016} obtained VLT UVES spectra for a total of eight stars in HP1, while four stars in Terzan 4 were observed with Keck NIRSPEC by \citet{Origlia2004}. We compare their results with our values for elements in common in Table \ref{comparison}.
The comparison with \cite{Barbuy2016} is generally good, within 0.2 dex, with several outliers, although \cite{Barbuy2016} do not give their errors. The outliers are C and N, which are known to vary due to stellar evolution and/or MP (see below), so a variation is expected given the distinct stars sampled. The two Si abundances are identical and those for the other two well-determined and well-behaved $\alpha$ elements, Mg and Ca, are within 0.1 dex, which is reassuring. \cite{Barbuy2016} found that their Na abundance is subsolar, as is our almost identical value, although again this is an element affected by MP. They also found that, of their $\alpha$ elements, O, Mg and Si were more abundant than Ca and Ti. We find all four of our $\alpha$ elements to be almost identical. 

\citet{Ortolani2011} find HP1 is $\sim13.7$ Gyr old, and may be one of the oldest
GCs in the Galaxy. This was corroborated by \citet{Dias2016}. \citet{Kerber2019} revisit this issue, now using a very deep Gemini South GeMS CMD. They use the \cite{Barbuy2016} metallicity of $-1.06\pm0.10$ and [$\alpha$/Fe]=+0.4 to derive an age of 12.8$\pm$0.9 Gyr, again confirming that HP1 is an old GC.
Our metallicity is slightly lower while our $\alpha$ values are around 0.25 dex, some 0.15 dex lower than their value. The effect of this difference on the derived age is model dependent, and in any case within the uncertainties. \citet{Kim2002} computed isochrones with and without $\alpha$ enhancement, and found
that the turnoff ages were 8\% smaller when going from [$\alpha$/Fe] = 0 to +0.3. 
This would imply a higher age by 
about 4\% using our $\alpha$ abundance, everything else being the same. We note that the deep near-IR CMD from \citet{Cohen2018} shows a strong drooping BHB.
It is clear that HP 1 is indeed an old BGC.

The \citet{Origlia2004} C abundance for Terzan 4 is very similar to our value (with the above caveat), while all of their $\alpha$ abundances are 
0.16 - 0.52 dex higher than our values, generally
much larger than the combined errors (their errors are 0.1 - 0.2 dex for each individual star). We note that both samples are small and that one star in our sample has a Ca abundance that is very discrepant, causing a large uncertainty, and Ca indeed has the largest difference. We are unsure of the cause of our general $\alpha$ discrepancy with \citet{Origlia2004}. 

\subsubsection{Radial velocity}
Radial velocities are powerful membership criteria, provide insight into internal cluster dynamics, and are key ingredients to derive the cluster orbit and thus constrain its origin.
As for metallicity, the H10 catalog is still a main source of RVs for Galactic GCs and is generally cited even in very recent studies like that of \citet{Perez-Villegas2018, Perez-Villegas2020}, who use H10 velocities to derive cluster orbits and possible origins, and search for correlations with chemical properties. H10 RV values for our sample are more robust than metallicities since the observational requirements are less demanding, and all of our sample have been the subject of a variety of velocity studies, most of which are incorporated in the H10 mean value.  Nevertheless, there are several clusters where either H10 does not list a value or his value is significantly different from ours. A new compendium by \cite{Baumgardt2019} is now available and widely used and includes all of our sample.
Here we detail the results for each cluster and compare them to these primary sources as well as any other recent studies we are aware of.

{\bf Terzan 2:} 
The H10 RV value for this cluster is 109.0$\pm$15.0 km/s, while our mean value is 133.2$\pm$1.4 km/s. Other values include 144.6$\pm$0.8 \citep{Vasquez2018}, 128.96$\pm$1.18 \citep{Baumgardt2019}, and 130.49$\pm$2.29 km/s \citep{Geisler2021}.
Our value is in the middle of the other values and in good  agreement with them, except for that of H10, although given his large uncertainty, the agreement is satisfactory. 

{\bf Terzan 4:}
The H10 RV value for this cluster is -50.0$\pm$2.9 km/s, while our mean value is -48.3$\pm$3.5 km/s. The only other value is -39.93$\pm$3.76 \citep{Baumgardt2019}.
Our value lies between the other values and is in good  agreement with them, especially H10.

{\bf HP1:}
The H10 RV value for this cluster is 45.8$\pm$0.7 km/s, while our mean value is 39.8$\pm$4.0 km/s. Other values include 60$\pm$13 \citep{Minniti1995b}, 45.8 \citep{Barbuy2006}, 40.0$\pm$0.5 \citep{Barbuy2016}, and 40.61$\pm$1.29 \citep{Baumgardt2019} km/s.
Our value is slightly below other values but in good  agreement with them, except for that of Minniti, which has a large uncertainty.

{\bf Terzan 9:}
The H10 RV value for this cluster is 59.0$\pm$10.0 km/s, while our mean value is 68.1$\pm$4.3 km/s. Other values include 71.4$\pm$0.4 \citep{Vasquez2018}, 29.31$\pm$2.96 \citep{Baumgardt2019}, 58.1$\pm$1.1 \citep{Ernandes2019}, and 71.58$\pm$2.28 km/s \citep{Geisler2021}.
Our value falls within the range of the other values and is in good  agreement with them (except for the \cite{Baumgardt2019} value, which appears very low),  especially the CaT-derived values \citep{Vasquez2018,Geisler2021}.  

{\bf Djorg 2:}
H10 does not list an RV value for this cluster. Our mean value is -150.2$\pm$4.7 km/s. Other values include -150$\pm$28 \citep{Dias2016}, -159.9$\pm$0.5 \citep{Vasquez2018}, -148.05$\pm$1.38 \citep{Baumgardt2019}, and -162.45$\pm$9.14 km/s \citep{Geisler2021}.
Our value falls within the range of the other values and is in good  agreement with them. 

{\bf NGC 6540:}
The H10 RV value for this cluster is -17.7$\pm$1.4 km/s, while our mean value is -14.4$\pm$1.1 km/s. Other values include -17.98$\pm$0.84 \citep{Baumgardt2019}, and -22.07$\pm$1.32 km/s \citep{Geisler2021}.
Our value falls slightly below the other values but is in reasonable  agreement with them. 

{\bf NGC 6642:}
The H10 RV value for this cluster is -57.2$\pm$5.4 km/s, while our mean value is -56.1$\pm$1.1 km/s. Other values include -41$\pm$24 \citep{Minniti1995b},  -33.23$\pm$1.13 \citep{Baumgardt2019}, and -51.4$\pm$2.5 km/s \citep{Geisler2021}.
Our value falls within the range of the other values and is in good  agreement with them, except for the \cite{Baumgardt2019} value, which appears low.

\subsection{Multiple populations}
GCs, long regarded as prototypical simple stellar populations, are now known to host star-to-star variations in a variety of chemical elements.  More specifically, \citet{Carretta2009} showed that all Galactic GCs well-studied up to then have at least a spread (or anti-correlation) in the content of their light-elements O and
Na. In some cases also a Mg and Al spread is observed, which is metallicity dependent \citep[e.g.,][]{Ventura2016}. (The only confirmed exception amongst Galactic GCs so far is Ruprecht 106, where \citet{Villanova2013} found that their sample of nine stars share a homogeneous chemical composition in all elements studied, including these four.) 
This is the so-called MP phenomenon. 
These abundance anomalies have also been observed in old, massive extragalactic GCs in Fornax \citep{Letarte2006, Larsen2012} and the  Magellanic Clouds (e.g., \citealt{Mucciarelli2009}), as well as massive intermediate-age Cloud clusters \citep{Martocchia2019}.
However, the investigation of MP so far has been limited almost
exclusively to non-GB, non-metal-rich GCs. \citet{Meszaros2020} investigated the APOGEE-2 data for 44 GCs, of which, after eliminating clusters with a small number of members as well as those with high reddening, only two considered Main Bulge GCs by \citet{Massari2019} remained, graphically illustrating the general lack of BGC studies in this key area.

Several theoretical models have been proposed to describe the formation and early evolution of GCs  to explain MP \citep[e.g.,][]{DAntona2016}.
The current most viable explanation involves a self-enrichment scenario, where subsequent generations of stars coexist in GCs that are formed from gas polluted by processed material produced by massive first generation stars \citep{Renzini2015}. A variety of possible
sources of polluters have been proposed: intermediate-mass asymptotic giant branch (AGB) stars \citep{DAntona2016}, fast-rotating massive stars \citep{Decressin2007},  massive binaries \citep{deMink2009} or supermassive stars \citep{Gieles2018}.  
Unfortunately, none of the currently proposed scenarios are able to account for the totality of MP phenomena now known \citep{Renzini2015, Bastian2018} and a successful formation scenario still eludes us. 


Generally, the spectroscopic investigation of  MP among the light elements only take into account elements heavier than N since it is very difficult to measure C and N in optical spectra. Fortunately, the APOGEE near-IR spectra are rich in CO, CN, and OH lines, allowing us to derive individual C, N, and O abundances for all our stars, after properly accounting for molecular equilibrium. However, one needs to be careful interpreting C and N abundances since they can be altered by the evolutionary status of the stars. Since in our investigation of MP we are interested only in differential abundances, our main concern in this regard is if our targets are pre- or post-RGB-bump stars, because evolutionary mixing effects that could produce differential effects generally appear at the bump. At that stage, [C/Fe] drops by about 0.4 dex and [N/Fe] increases by about 0.6 dex with respect to pre-bump values, while the [C/Fe] and [N/Fe] values of only pre- or only post-RGB-bump stars are approximately constant, independent of the evolutionary stage \citep{Gratton2000}.
We checked the evolutionary stage of our targets by comparing their
T$_{\rm eff}$ to the 
temperature of the RGB-bump taken from 12.5 Gyr isochrones calculated in the metallicity range from [Fe/H]=-1.4 to [Fe/H]=-0.8, appropriate for our sample. Isochrones were obtained from the Padova database \citep{Bressan2012}. 
The bump values we find are between 4800 K and 4914 K, depending on the metallicity. 
We find that all our stars are cooler than the bump, and therefore post-bump stars, with the exception of a couple of targets that lie in this range. We conclude that our sample is not significantly affected by evolutionary effects in its C and N abundances, at least differentially, as virtually all stars lie in the same evolutionary phase.
We also find that the N abundance range is very large amongst our targets and that N is very effective in separating 1G and 2G stars. 

Our study of MP is unfortunately limited by several factors. First, we generally only have quite small samples in our clusters, with only a single to a few 1G and 2G stars per GC to trace MP. Thus, the statistical significance of any trends is not as high as originally intended, given our initial goals. We do note that our preliminary examination of the final CAPOS data, which will be available in DR17, indicates that most if not all of our clusters will end up with more members with good data, so the statistics will improve.
Second, we have argued that abundances for at least Fe, and probably many other elements studied here, if not all,  in 2G stars are not well measured by ASPCAP. This of course constrains our ability to measure robust trends. Also, as noted above, the evolutionary state of a star causes surface abundance variations caused by mixing and not intrinsic variations, although these may be relatively small for our sample.
Given these limitations, we
simply assume that ASPCAP abundances of all elements for all stars, including 2G, are correct but only carryout a qualitative analysis. In subsequent papers we will derive independent atmospheric parameters and abundances for all stars and perform a more detailed, quantitative study. Also, in order to arrive at more statistically significant sample sizes than possible with the small number of stars in some GCs, we bin our GCs into two metallicity groups. Given their mean [Fe/H] values, it was
natural to divide  them into a metal-rich  group that  includes Terzan 2, HP 1, Djorg 2, NGC 6540, and NGC 6642, with  mean metallicity ranging from  [Fe/H]=-0.85 to -1.2, and a total of 28 stars, and a  metal-poor group with Terzan 4 and Terzan 9, which both have [Fe/H]=-1.4,
yielding a total of 12 stars. 
All clusters in the same metallicity bin generally follow similar distributions in all elements, allowing us to better explore trends by combining them.

\begin{figure*}
\centering
   \includegraphics[width=16cm]{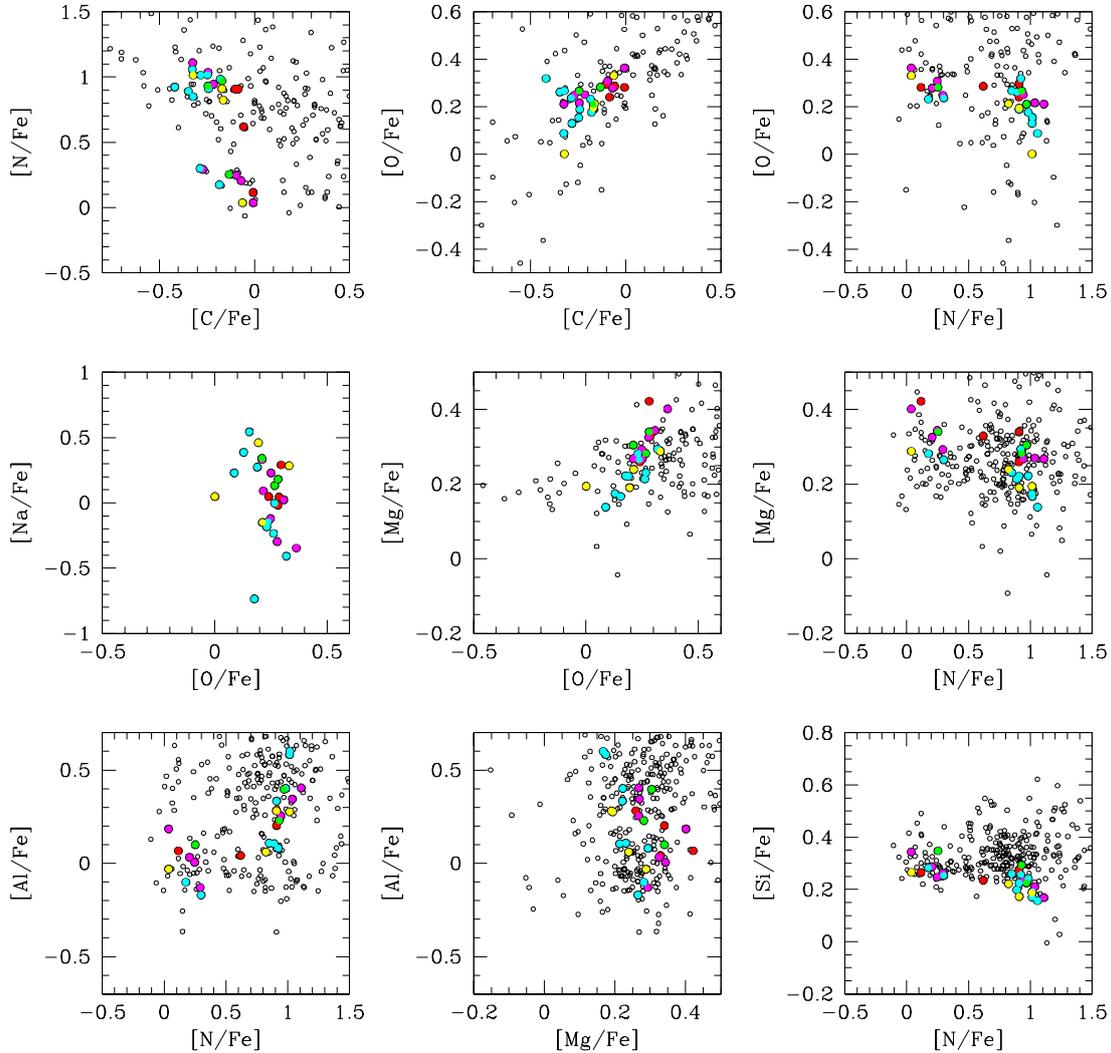}
      \caption{Light-element (anti)correlations for our higher metallicity clusters: Terzan 2 (red circles),  
      Djorg 2 (magenta), HP1 (cyan), NGC 6540 (yellow), and NGC 6642 (green). Comparison objects (empty circles) from \citet{Masseron2019} and \citet{Meszaros2020} were selected in the metallicity range [Fe/H]$>-1.3$. Typical error bars are about 0.1 dex per element.}
    \label{anti_m08_m12}
\end{figure*}

\begin{figure*}
\centering

   \includegraphics[width=17cm]{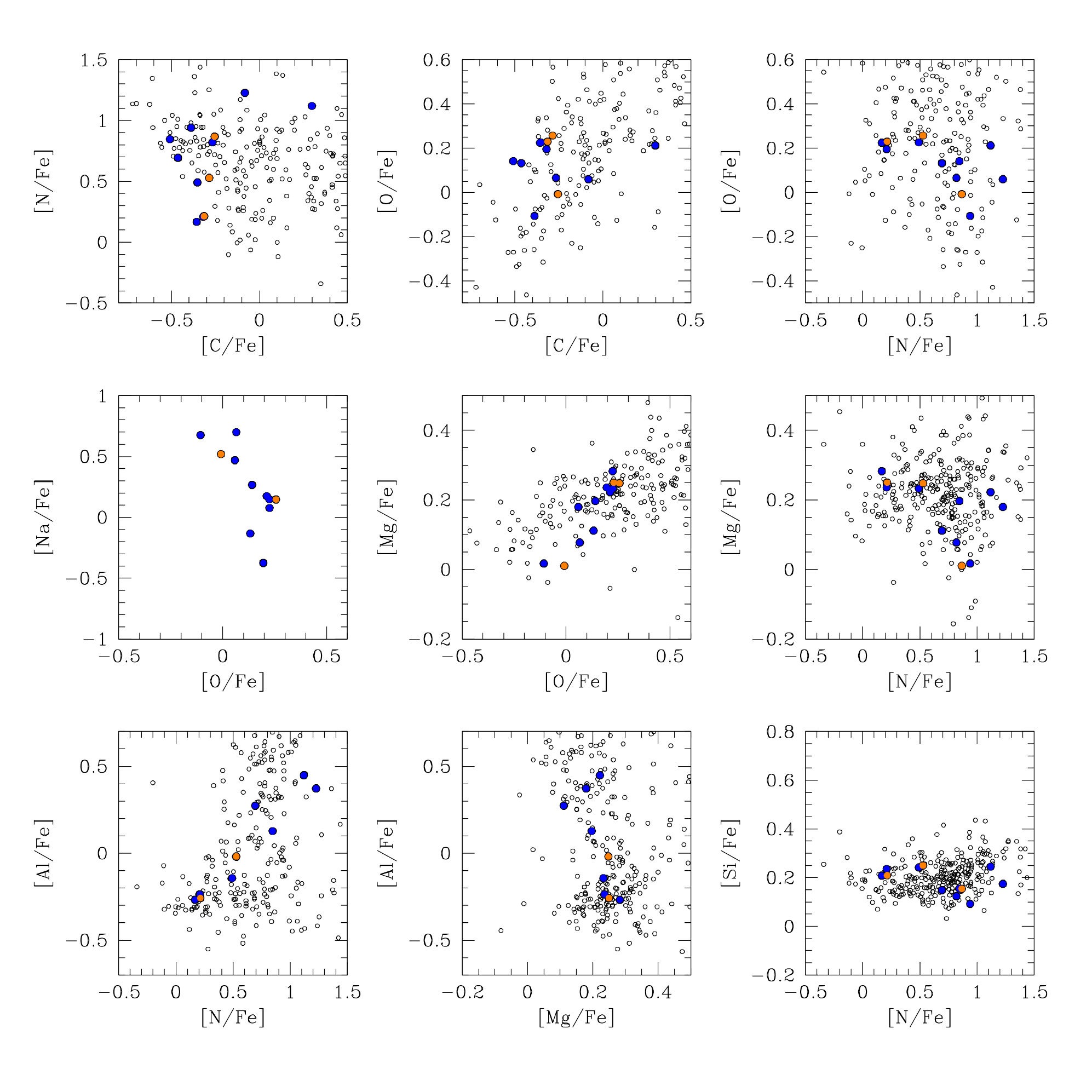}
      \caption{Light-element (anti)correlations for our lower metallicity clusters
       Terzan 9 (blue circles) and Terzan 4 (orange circles). Comparison objects (empty circles) from \citet{Masseron2019} and \cite{Meszaros2020} were selected in the metallicity range -1.6$\leq [Fe/H] \leq $ -1.2. Typical error bars are about 0.1 dex per element.}
    \label{anti_m14}
\end{figure*}

In Figs. \ref{anti_m08_m12} and \ref{anti_m14} we compare distributions for a variety of elements up to Si for each group. Stars of each cluster are represented by filled circles of a given color, while comparison GC stars from the literature \citep{Masseron2019} are represented by open circles. For each metallicity range, the comparison sample was selected to fall in the [Fe/H] interval covered by our BGCs, so that each group is compared with clusters of similar metallicity.
These comparison clusters were studied using the same instrument, but analyzed using atmospheric parameters independent from ASPCAP, which in general will yield different results for 2G stars. 
Typical internal errors are $\sim$0.05 dex in each element.

The relatively metal-rich (Fig. \ref{anti_m08_m12}) and metal-poor (Fig. \ref{anti_m14}) clusters each display a wide variety of interesting (anti)correlations with general similarities, but also some important differences between them. Correlations or anticorrelations appear in most of the plots. In general, the distribution of our metal-rich sample is tighter than that of the metal-poor clusters, even though the former includes more than twice the sample of the latter. Also, both cluster samples generally follow the trend of previous GC studies. Clearly, all of our BGCs exhibit MP phenomena.

N is a key element for detecting and delineating MP. All five metal-rich clusters appear to be essentially bimodal in N, with only 1G and 2G stars, with the lone exception of an intermediate star in Terzan 2. On the other hand,  while the metal-poor sample clearly have a very wide range in N abundance, they do not show a clear N bimodality, with 3 of the 12 stars lying at intermediate [N/Fe]. There are only two clusters in the \citet{Meszaros2020} APOGEE sample of 31 GCs in which N bimodality can be convincingly determined: NGC 288 and M10, with metallicities of -1.18 and -1.35, falling essentially between our two metallicity groups.

Both metallicity samples have clear and tight Mg versus O correlations, and relatively clear and tight O versus N, Na versus O, Mg versus N, Al versus Mg, and Si versus N anticorrelations. Metal-rich clusters also show a tight O versus C correlation and relatively tight N versus C anticorrelation, while the metal-poor sample do not show clear (anti)correlations in these plots. Thus, metal-rich, and to a lesser extent metal-poor, 1G N-poor stars ([N/Fe]$\leq $+0.3) are also generally enhanced in C and O compared to 2G stars, and to a lesser degree in Mg and Si as well. This  fact points toward  Si-leakage of the Mg-Al chain inside the polluters that are responsible for the chemical pattern of the 2G stars. 

As far as C and O are concerned, both elements show large spreads, that are anticorrelated with N for the metal-rich sample, while the metal-poor sample shows significant spreads in both elements but not as clear anticorrelations, especially for the two most N-rich stars in Terzan 9, which deviate from the general trends. Mg is strongly correlated with O in both samples, with very similar slopes that are steeper than that shown by the comparison GCs.  Mg appears to be  anticorrelated with N in both samples, with the same two exceptional Terzan 9 N-rich stars, which also stand out from the general, albeit very weak, Si:N anticorrelation.

Al is another element strongly associated with MP. Indeed, the \citet{Meszaros2020} definition of 2G stars is [Al/Fe]$>$+0.3, while we are using [N/Fe]$>$+0.7. In the Al:N plots, we find that the metal-poor stars show a clear correlation and all stars would be classified as 2G using either element classification, except for one star that qualifies as 2G from its N abundance but not from Al. 
However, for the metal-rich sample, although the general tendency is similar, there is a wider variety of behavior, with one ``Al 2G'' star having a very low N abundance and a group of ``N 2G'' stars having low Al abundance. Both comparison samples show similar distributions and examples of a few ambiguous cases.
Clearly, for the purpose of separating bonafide 1G and 2G stars, errors should also be taken into account. In addition, using just one element is not ideal. The combination of two elements, e.g. N and Al together, is preferable.
In this study, we associate the existence of MP mainly with the presence of  intrinsic spreads in C, N, Na, and Al.

Meszaros et al. also investigate the Al:Mg distribution. Although their anticorrelation is often associated with MP, along with the Na:O anticorrelation, not all GCs show a Al:Mg anticorrelation. This is at least in part due to the fact that the Mg-Al cycle cannot start in high metallicity GCs since their interiors do not reach the necessary high temperatures to activate it \citep{Meszaros2020}. Indeed, they find that it is hard to generalize MPs from the properties of Al versus Mg, and in reality every cluster has its own specific pattern of MPs showing a wide variety. The behavior of other, non-CAPOS GCs is seen in the figures, where a mild but still visible anticorrelation between Al and Mg exists but with large scatter.

We agree with the Meszaros et al. general finding that an anticorrelation between Al and Mg is weakly present, in both our cluster groups, with a typical Mg range of $\sim$0.2 dex, much smaller than that of Al, which varies by about 0.8 dex. The agreement between the distribution of our samples and the comparison stars is good.

The archetypical MP Na versus O anticorrelation, exhibited by almost all GCs in the Galaxy \citep{Carretta2009}, is present but weak for the metal-rich BGCs and stronger in the metal-poor BGCs, especially Terzan 9. The metal-rich clusters show a clear Na spread but are accompanied by only a moderate 
O variation. The two metal-rich GCs with the largest number of stars - HP 1 and Djorg 2 - demonstrate the anticorrelation is indeed present within each cluster, while the smaller samples of the other metal-rich BGCs do not. However, that a real variation is present in O for these clusters is borne out by their behavior in the O:C, O:N, and Mg:O plots, displaying clear (anti)correlations, especially for NGC 6540.

This behavior for metal-rich BGCs to exhibit a weak or even negligible O spread was already noticed by \citet{Munoz2017, Munoz2018, Munoz2020} where the authors found that in the BGCs NGC~6440, NGC~6528, and NGC~6553, O does not vary substantially between 1G and 2G stars, despite an obvious Na spread.  However, we note that the metallicity of these metal-rich BGCs is  significantly higher than that of our metal-rich sample, ranging from -0.14 to -0.50.
Table \ref{Element} in fact reveals that O has a low spread in all GCs from our sample, while Na shows a significant spread in most clusters with a large sample size. 
The comparison clusters do
not have any information about Na as \citet{Masseron2019} do not derive abundances for this element due to low reliability for low metallicity stars.

\subsection{Fe-peak elements}

Very few measurements of Fe-peak elements other than Fe are available for individual stars of BGCs. In particular, no such measurements exist for any of our sample. ASPCAP produces abundances of the Fe-peak elements V, Cr, Mn, Co, Ni, and Cu. However, the reliability of these abundances can vary with the number of lines, their strengths, how well they compare to previous studies, limitations of the pipeline, etc. \citet{Henrik2020} have undertaken a painstaking assessment of the reliability of ASPCAP abundances. They rate Mn and Ni as the best Fe-peak species in both accuracy and precision, and find problems with V, Cr, Co, and Cu. Therefore, we abide by their assessment and only investigate these two elements. All Fe-peak elements will be investigated by deriving independent atmospheric parameters for the same APOGEE spectra, which we are pursuing.

\begin{figure*}
\centering
   \includegraphics[width=17cm]{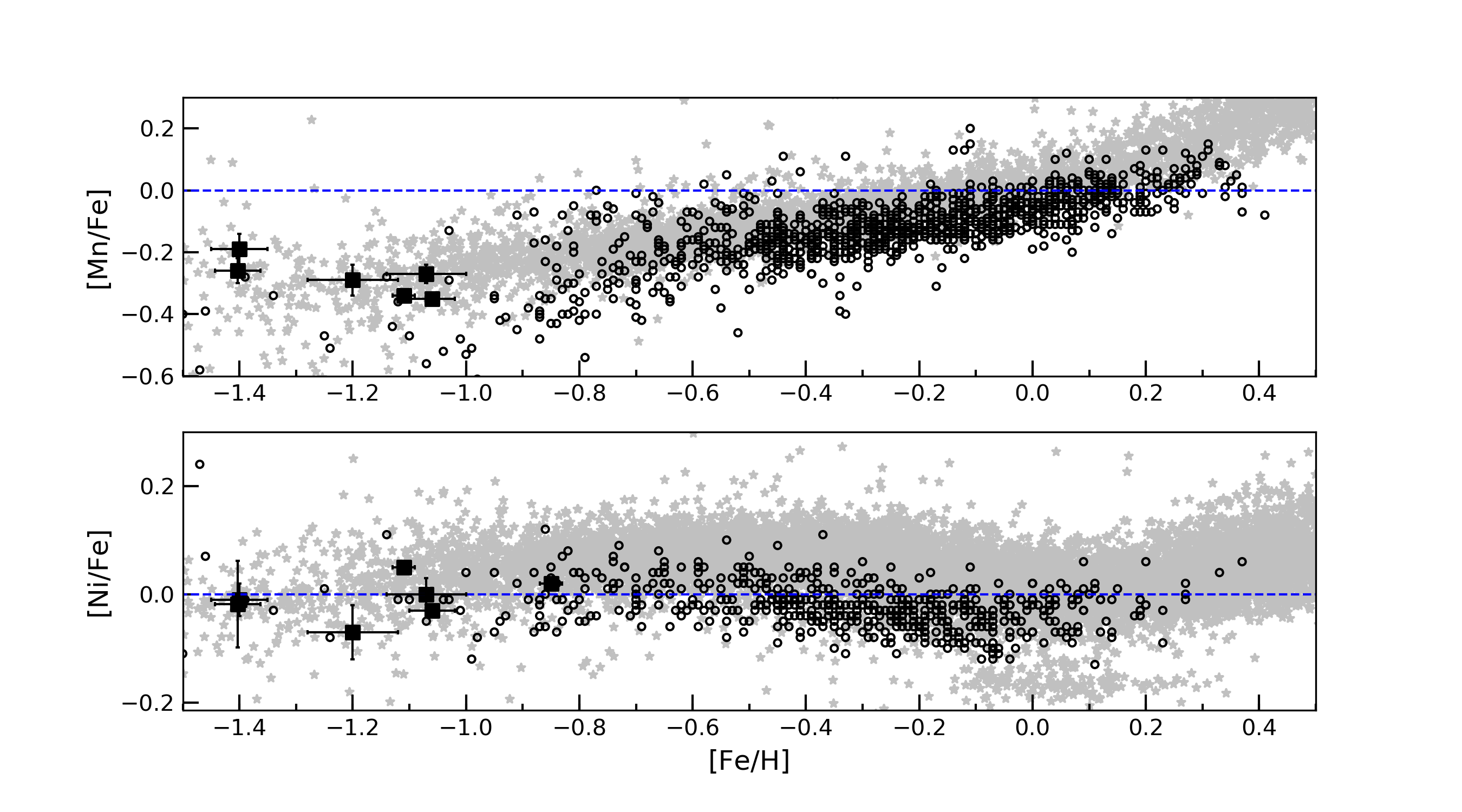}
      \caption{Mean abundance ratio of the Fe-peak elements Mn (top) and Ni (bottom) for each of our clusters (filled squares with error bars), compared with 
APOGEE bulge stars (asterisks - Rojas-Arriagada et al. 2020) and the general trend of disk/halo stars \citep[circles-][]{Reddy2003, Reddy2006, Battistini2015, Lomaeva2019}. Our BGCs generally follow the bulge field-star trend.
      }
    \label{f_iron}
\end{figure*}

The behavior of the mean [Mn/Fe] and [Ni/Fe] versus [Fe/H] in our clusters is shown in Fig.~\ref{f_iron}, and compared with the trends of APOGEE (ASPCAP) bulge field stars \citep{Rojas-Arriagada2020}, as well as disk and halo field stars \citep{Reddy2003, Reddy2006, Battistini2015, Lomaeva2019}, whose abundances are derived from optical spectra. First, notably, our sample covers a metallicity range so far relatively underexplored in both species in all three Galactic components. Indeed, the fraction of bulge field stars with metallicities below $\sim$-1 is very small \citep{Rojas-Arriagada2020}, and although disk stars have all but disappeared at these low metallicities while the halo is prominent, these elements are poorly studied. However, our results lie well within the bulge field star locus at these low metallicities, so there is good general agreement between our bulge cluster and field star samples.

However, the mean bulge [Mn/Fe] field star abundances from APOGEE show a small systematic offset over the full metallicity range from  those of disk/halo stars derived from optical spectra , with the APOGEE [Mn/Fe] being about 0.1 dex more enhanced. We note that this is very similar to the zero-point shift of +0.121 dex applied to the raw ASPCAP value to make giants with solar ASPCAP [M/H] have [Mn/Fe]= 0. \citet{Henrik2020} note that the accuracy of their comparison of APOGEE to optical results varies substantially depending on the comparison sample used, suggesting that much of the differences seen may well arise from problems with the optical samples. At the metallicities of our BGCs, the offset appears to increase although the sample sizes are much smaller than at higher metallicities. The small Mn abundance ratio offset between the bulge and the disk is very similar for [Fe/H]$>-1.0$ \citep{Lomaeva2019}, with a steady increase in [Mn/Fe] with metallicity in both samples due to the strong dependence of Mn yields on metallicity in all production sites \citep{Woosley1995, Woosley2011}. However, below [Fe/H]=$-1$, field halo stars form a plateau at about [Mn/Fe]$\approx -0.4$ \citep{Nissen2000, Adibekyan12, Reggiani2017}, while our results indicate a nearly constant but somewhat higher ratio ([Mn/Fe]$\approx -0.25$) in BGCs and field stars. In fact, our BGCs fall near or even above the upper limit of the halo field star distribution, especially for our two most metal-poor GCs. This suggests that Mn production in the early stages of the bulge evolution was generally more efficient than in the halo.
The GB chemically evolved extremely fast \citep[e.g.,][]{Bidin2021}, and [Fe/H]$<-1.0$ should correspond to the first 0.5~Gyr of evolution \citep{Lian2020}. Hence, a difference in [Mn/Fe] should have risen in the very first stages when, according to recent models, hypernovae must have played a key role. For example, a lower frequency of hypernovae in the bulge with respect to the disk/halo could have caused an increased production of Mn in the bulge \citep[][see their Fig.5]{Grimmett2020}.

\citet{Henrik2020} find Ni to be even more accurate and precise than Mn, with a zero-point shift applied to raw ASPCAP values for giants of only -0.016 dex. They do note that there is an issue with ASPCAP that produces a thin horizontal sequence at [Ni/Fe]$\sim$-0.2, which only affects giants with $T_{eff}<4000K$ and can be seen in Fig.~\ref{f_iron} around solar metallicity. Only two of our sample lie in this temperature range, the two 1G stars in Terzan 2, and their Ni abundance is consistent with other bulge stars at this metallicity. 

We find our BGCs to lie within the range of the Ni abundances of similar-metallicity bulge field stars, although four of our five more metal-rich clusters generally lie to lower [Ni/Fe] values than the mean of the field star distribution. 
There are very few halo stars in the metallicity range of our BGCs, but the values overlap. However, again there is a clear offset between the bulge and disk/halo field comparison samples at higher metallicities, with the bulge about 0.1 dex enhanced. Otherwise, the two samples have very similar shapes with metallicity. This enhancement in [Ni/Fe] observed at higher metallicities between the bulge and disk/halo field stars \citep{Lomaeva2019} does not clearly persist in the metal-poor range of our BGCs, where of course sample sizes are minimal. However, in our clusters, which are more precise than single field stars, Ni seems to flatten at about solar value, while Ni could flatten at -0.1 in the halo \citep[see, e.g., Fig.16 of ][]{Bensby2014} and \citep[Fig.8 of][]{Adibekyan12}. The data are too scarce to say anything more detailed at this point, but if this offset was confirmed, the bulge-disk/halo offset would in fact still remain even in this metallicity range. An offset that is constant down to [Fe/H]=-1.4 is what we find for Mn, so finding the same behavior for Ni would not be too surprising.

\section{Conclusions}

We present an overview and initial ASPCAP results of CAPOS, the 
bulge Cluster APOgee Survey. This survey is designed to obtain detailed abundances and kinematics for as complete a sample of bona fide BGCs as possible using the unique advantages of APOGEE in order to exploit their extraordinary Galactic archaeology attributes. We aim to help gather the first definitive database on the BGC system, 
provide accurate chemical characterizations of some of the oldest objects, and perhaps the oldest, in the Milky Way,
study the very complex nature of the GB and uncover any underlying correlations, determine their orbits, and search for MPs in the BGC system and compare them to their halo (and disk) counterparts. CAPOS observed 16 cataloged BGCs, and also several BGC candidates recently uncovered in the same fields, as well as a large number of field stars selected from the general field, the 
K2 Galactic Archaeology Program (K2GAP), and the EMBLA and PIGS surveys for metal-poor stars.

Here we present initial CAPOS results for seven cataloged BGCs that were analyzed using the APOGEE pipeline ASPCAP as part of the 16th data release of the SDSS IV survey. ASPCAP provides atmospheric parameters, an accurate RV  and detailed abundances for  some 20 chemical elements. In future papers, we will derive abundances for this sample from boutique analyses, including stars with S/N below the ASPCAP limit, investigate the non-BGC M22, as well as results for the CAPOS field stars observed in DR16.  Subsequent papers will include additional BGCs and field stars obtained after the data included in DR16. We find no clustering in metallicity:RV space for our 9 targets in the BGC candidate Minni 51 and thus no evidence for its reality.

The observed GC giant targets were carefully selected based on spatial position, existing photometry, radial velocity, proper motion
and any spectroscopic metallicity information available in order to maximize cluster membership. Our selection procedure yielded very high membership probability for our final sample of 40 giants in the seven clusters, with three to ten members per cluster having S/N$>$70. This is sufficient to derive good mean abundances and velocities and explore MP.

We detect a small but significant systematic correlation between the ASPCAP metallicity and effective temperature. We also find a trend within each of our clusters for stars with the highest N abundances (2G stars) to also show a higher metallicity than their 1G counterparts. We interpret this as due to a known issue with ASPCAP to overestimate effective temperatures for 2G stars. We used the metallicity of 1G stars to correct [Fe/H] values for 2G stars, and used these corrected values together with those of 1G stars to derive the mean [Fe/H] for our clusters. 
We derive mean [Fe/H] values of 
-0.85$\pm${ 0.04} for Terzan 2 from four stars, 
-1.40$\pm${ 0.05} for Terzan 4 from three stars, 
-1.20$\pm${ 0.10} for HP 1 from ten stars, 
-1.40$\pm${ 0.07} for Terzan 9 from nine stars, 
-1.07$\pm${ 0.09} for Djorg 2 from seven stars, 
-1.06$\pm${ 0.06} for NGC 6540 from four stars, and 
-1.11$\pm${ 0.04} for NGC 6642 from three stars, 
where the error is the standard deviation.
All of our clusters lie well below the most metal-poor peak of the bulge field star distribution of \citet{Rojas-Arriagada2020}.

We also use the ASPCAP values for 1G stars to determine mean abundances for 11 other elements plus the mean global-fit [$\alpha$/Fe] abundance (our best proxy for [$\alpha$/Fe]). The latter we derive to about 0.02 dex. We also derive mean cluster radial velocities to typically a few km/s. We compared our values for [Fe/H], [$\alpha$/Fe], and RV with those in the literature, and find generally good agreement. We believe our values are the most precise and accurate available. Only two of our clusters have been observed previously with high-resolution spectroscopy, and agreement with individual element abundances is generally good. We corroborate and reinforce the finding of \citet{Horta2020} that purported Main Bulge and
Main Disk GCs, presumably formed in situ, have [Si/Fe] abundances slightly higher than their presumably accreted counterparts at the same metallicity, associated with various accretion events. Including our CAPOS clusters 
significantly increases the sample of well studied Main Bulge GCs and also extends them to lower metallicity.

An important goal of CAPOS is to investigate MP, as GCs are known to host star-to-star variations, but the study of MP in BGCs is severely lacking in comparison to their disk and  halo counterparts.
Unfortunately, the small number of stars per cluster and problems with ASPCAP abundances for 2G stars prevent us from carrying out more than a qualitative assessment of MP in our sample.
To explore the possible dependence of MP behavior on metallicity and increase the sample size, we binned our GCs in two metallicity groups: a metal-richer group 
including  Terzan 2, Djorg 2, HP1, NGC 6540, and NGC 6642, 
with metallicity ranging from -0.85 to -1.2; and the metal-poorer clusters Terzan 9 and Terzan 4, with  metallicity [Fe/H]=-1.4. 
We find clear MP within each cluster and element, and explored their behavior in each group, which generally follow the trend of previous studies of GCs of similar metallicity. 
All five metal-rich clusters appear to be essentially bimodal in N, while the metal-poor GCs do not. Both metallicity samples have clear and tight Mg versus O correlations and relatively clear and tight O versus N, Na versus O, Mg versus N, Al versus Mg, and weak Si versus N anticorrelations. Metal-rich clusters also show a tight O versus C correlation and relatively tight N versus C anticorrelation, while the metal-poor sample do not exhibit clear (anti)correlations in these plots.  Metal-poor Al-rich stars are also N-rich, while this is not always the case for the metal-rich sample. An anticorrelation between Al and Mg is weakly present in both our
cluster groups, with a typical Mg range much smaller than that of Al.
The archetypical anticorrelation of Na:O is clearly detected for the metal-poor group but only weakly present in the metal-rich group, which is characterized more by a strong Na variation and only a small O variation. 

We finally explore the abundances of the most reliable ASPCAP Fe-peak elements Mn and Ni, and compare  their trends with the Galactic disk, halo, and bulge field stellar populations. We find that the abundances of these elements in general follow the trends observed for bulge stars. 
Below a metallicity of -1, field halo stars form a plateau at a [Mn/Fe] abundance slightly higher than that of BGCs and field stars in this regime. Our
BGCs fall near or even above the upper limit of the halo field star distribution, especially for our two most metal-poor GCs. This suggests that Mn production in the early stages of the bulge evolution was generally more efficient than in the halo.
Our BGCs lie within the range of the Ni abundances of similar metallicity bulge field stars, although generally to lower
[Ni/Fe] values than the mean of the field star distribution, 
and overlap with  halo stars in the metallicity range of our BGCs.

We are currently analyzing more CAPOS GC data, as well as that for field stars, and  look forward to the DR17 release (scheduled for December 2021) of additional very recent observations, all of which will further contribute to this legacy database. A second CAPOS paper 
graphically illustrates the potential of 
this project, in this case exploring
the intriguing BGC FSR 1758 \citep{Romero-Colmenares2021}. 
Finally, we are enthusiastic that the CAPOS goals to further investigate BGCs will continue to be carried out as part of the SDSS-V Open Fiber program with the Milky Way Mapper using APOGEE.

\begin{acknowledgements}
This paper is dedicated to the memory of George Wallerstein. George was a true pioneer in the field of detailed abundance studies of globular clusters, 
as well as making many other stellar contributions. He was the founding father of the Astronomy Department of the University of Washington, where he taught many generations of future astronomers, and was the thesis advisor of both the first author and V.S.
We would like to warmly acknowledge SDSS-IV and especially all of those who have contributed to the outstanding success of both APOGEE North and South, including M. Skrutskie and J. Wilson for the original idea, S. Majewski and K. Cunha for the original science motivation, Majewski, Skrutskie and Wilson for putting together and carrying out the project, V. Smith, K. Cunha and C. Allende-Prieto for helping identify species and their transitions, and Allende-Prieto, J. Holtzman and D. Nidever for developing the ASPCAP pipeline. We are deeply grateful to all those responsible for carrying out all phases of the CAPOS observations, especially the observation of the last field, which was the only bulge field observed by APOGEE-2S during all of 2019.

We thank David Nataf, Louise Howes and Martin Asplund, who provided us access to the EMBLA sources.
D.G., S.V., J.E.O., M.C., D.M., M.Z., J.A.-G., F.M. and R.R.M. gratefully acknowledge support from the Chilean Centro de Excelencia en Astrof\'isica y Tecnolog\'ias Afines (CATA) BASAL grant AFB-170002.
D.G. also acknowledges financial support from the Dirección de Investigación y Desarrollo de
la Universidad de La Serena through the Programa de Incentivo a la Investigación de
Académicos (PIA-DIDULS), as well as the moral and amorous support of M.E. Barraza. SV gratefully acknowledges the support provided by FONDECYT Regular No. 1170518.
J.G.F-T is supported by FONDECYT No. 3180210 and Becas Iberoam\'erica Investigador 2019, Banco Santander Chile.
D.M. is also supported  by the ANID - Millennium Science Initiative Project ICN12\_009,
awarded to the Millenium Institute of Astrophysics (MAS) and by Proyecto FONDECYT No. 1170121.
Support for M.C., D.M., M.Z., J.A.-G, and F.M. is also provided by MAS and by FONDECYT projects \#1171273 and \#1191505.
C.M.  thanks the support  provided by ANID  through  Beca Postdoctorado en el Extranjero convocatoria 2018 folio 74190045.
C.M. thanks the support provided by  FONDECYT Postdoctorado No.3210144.
Financial support for C.C. is provided by Proyecto FONDECYT Iniciaci\'on a la Investigaci\'on 11150768 and MAS.
B.T. gratefully acknowledges support from the National Natural Science Foundation of China under grant No. U1931102 and support from the Hundred-Talent Project of Sun Yat-sen University.
J.A.-G. acknowledges support by Proyecto Fondecyt Regular 1201490. A.M. and C.M. acknowledge support by FONDECYT Regular 1181797.
C. E. F. L. acknowledges a postdoctoral fellowship from the CNPq and to MCTIC/FINEP (CT-INFRA grant 0112052700) and the EmbraceSpace Weather Program for the computing facilities at INPE.
DGD acknowledges support from call N.785 of 2017 of the Colombian Departamento Administrativo de Ciencia, Tecnolog\'ia e Innovaci\'on, COLCIENCIAS.
AA and ES gratefully acknowledge funding by the Emmy Noether program from the Deutsche Forschungsgemeinschaft (DFG). NFM gratefully acknowledges support from the French National Research Agency (ANR) funded project ``Pristine'' (ANR-18-CE31-0017) along with funding from CNRS/INSU through the Programme National Galaxies et Cosmologie and through the CNRS grant PICS07708. AA, ES and NFM thank the International Space Science Institute, Bern, Switzerland for providing financial support and meeting facilities to the international team ``Pristine''. 
SzM has been supported by the J{\'a}nos Bolyai Research Scholarship of the Hungarian Academy of
Sciences, by the Hungarian NKFI Grants K-119517 of the Hungarian National
Research, Development and Innovation Office, and by the {\'U}NKP-20-4 New National Excellence Program of the
Ministry for Innovation and Technology.
Funding for LH was provided by the Chilean National Agency for Research and Development
(ANID)/CONICYT-PFCHA/DOCTORADO NACIONAL/2017-21171231 grant.
T.C.B. acknowledges partial support from grant PHY 14-30152, Physics Frontier Center/JINA Center for the Evolution of the Elements (JINA-CEE), awarded by the U.S. National Science Foundation.
R.R.M. also acknowledges partial support from  FONDECYT project N$^{\circ}1170364$.
C.M. also acknowledges  the support provided by  FONDECYT No. 1181797. 
F.G. acknowledge support by CONICYT-PCHA Doctorado Nacional 2017-21171485.
DAGH acknowledges support from the State Research Agency (AEI) of the
Spanish Ministry of Science, Innovation and Universities (MCIU) and the
European Regional Development Fund (FEDER) under grant AYA2017-88254-P. 
 We also thank the referee for useful comments that improved the paper.
\\\\
Funding for the Sloan Digital Sky Survey IV has been provided by the Alfred P. Sloan Foundation, the U.S. Department of Energy Office of Science, and the Participating Institutions. SDSS-IV acknowledges
support and resources from the Center for High-Performance Computing at
the University of Utah. The SDSS web site is www.sdss.org.
\\\\
SDSS-IV is managed by the Astrophysical Research Consortium for the 
Participating Institutions of the SDSS Collaboration including the 
Brazilian Participation Group, the Carnegie Institution for Science, 
Carnegie Mellon University, the Chilean Participation Group, the French Participation Group, Harvard-Smithsonian Center for Astrophysics, 
Instituto de Astrof\'isica de Canarias, The Johns Hopkins University, 
Kavli Institute for the Physics and Mathematics of the Universe (IPMU) / 
University of Tokyo, Lawrence Berkeley National Laboratory, 
Leibniz Institut f\"ur Astrophysik Potsdam (AIP),  
Max-Planck-Institut f\"ur Astronomie (MPIA Heidelberg), 
Max-Planck-Institut f\"ur Astrophysik (MPA Garching), 
Max-Planck-Institut f\"ur Extraterrestrische Physik (MPE), 
National Astronomical Observatories of China, New Mexico State University, 
New York University, University of Notre Dame, 
Observat\'ario Nacional / MCTI, The Ohio State University, 
Pennsylvania State University, Shanghai Astronomical Observatory, 
United Kingdom Participation Group,
Universidad Nacional Aut\'onoma de M\'exico, University of Arizona, 
University of Colorado Boulder, University of Oxford, University of Portsmouth, 
University of Utah, University of Virginia, University of Washington, University of Wisconsin, 
Vanderbilt University, and Yale University.
\\\\
This work has made use of data from the European Space Agency (ESA) mission {\it Gaia} (\url{https://www.cosmos.esa.int/gaia}), processed by the {\it Gaia} Data Processing and Analysis Consortium (DPAC, \url{https://www.cosmos.esa.int/web/gaia/dpac/consortium}). Funding for the DPAC has been provided by national institutions, in particular the institutions participating in the {\it Gaia} Multilateral Agreement.
\\\\
Based on observations obtained through the Chilean National Telescope Allocation Committee through programs  CN2017B-37, CN2018A-20, CN2018B-46, CN2019A-98 and CN2019B-31.
Partly based on observations obtained with MegaPrime/MegaCam, a joint project of CFHT and CEA/DAPNIA, at the Canada-France-Hawaii Telescope (CFHT) which is operated by the National Research Council (NRC) of Canada, the Institut National des Science de l'Univers of the Centre National de la Recherche Scientifique (CNRS) of France, and the University of Hawaii.

We gratefully acknowledge the use of data from the ESO Public Survey program ID 179.B-2002 taken with the VISTA telescope and data products from the Cambridge Astronomical Survey Unit.
\end{acknowledgements}


\end{document}